\documentclass[conference]{IEEEtran}
\IEEEoverridecommandlockouts
\usepackage{cite}
\usepackage{amsmath,amssymb,amsfonts}
\usepackage{algorithmic}
\usepackage{graphicx}
\usepackage{textcomp}
\usepackage{xcolor}
\usepackage{subcaption}
\usepackage{listings}
\usepackage{enumitem}
\usepackage[ruled]{algorithm2e}
\usepackage{todonotes}
\usepackage{caption}
\usepackage{booktabs}
\usepackage{tagging}
\usepackage{sc24repro}
\usepackage{fancyhdr}

\usetag{explanation}
\usetag{example}

\DeclareCaptionFont{8pt}{\fontsize{8pt}{6pt}\selectfont}
\def\BibTeX{{\rm B\kern-.05em{\sc i\kern-.025em b}\kern-.08em
    T\kern-.1667em\lower.7ex\hbox{E}\kern-.125emX}}

\begin{document}

\pagestyle{fancy}
\fancyhead[CO,CE]{ACCEPTED AT THE INTERNATIONAL PARALLEL DATA SYSTEMS WORKSHOP, 2024}
\title{Exploring DAOS Interfaces and Performance}

\makeatletter 
\newcommand{\linebreakand}{%
  \end{@IEEEauthorhalign}
  \hfill\mbox{}\par
  \mbox{}\hfill\begin{@IEEEauthorhalign}
}
\makeatother 

\author{
\IEEEauthorblockN{Nicolau Manubens}
\IEEEauthorblockA{\textit{European Centre for Medium-Range} \\
\textit{Weather Forecasts (ECMWF)}\\
nicolau.manubens@ecmwf.int}
\and
\IEEEauthorblockN{Johann Lombardi}
\IEEEauthorblockA{\textit{DAOS Foundation}}
\and
\IEEEauthorblockN{Simon D. Smart}
\IEEEauthorblockA{\textit{ECMWF}}\\
\and
\IEEEauthorblockN{Emanuele Danovaro}
\IEEEauthorblockA{\textit{ECMWF}}\\
\linebreakand
\IEEEauthorblockN{Tiago Quintino}
\IEEEauthorblockA{\textit{ECMWF}}\\
\and
\IEEEauthorblockN{Dean Hildebrand}
\IEEEauthorblockA{\textit{Google}}
\and
\IEEEauthorblockN{Adrian Jackson}
\IEEEauthorblockA{\textit{EPCC, The University of}\\
\textit{Edinburgh}}
}
\IEEEaftertitletext{\vspace{-1\baselineskip}}

\maketitle
\thispagestyle{fancy}

\begin{abstract}

Distributed Asynchronous Object Store (DAOS) is a novel software-defined object store leveraging Non-Volatile Memory (NVM) devices, designed for high performance. It provides a number of interfaces for applications to undertake I/O, ranging from a native object storage API to a DAOS FUSE module for seamless compatibility with existing applications using POSIX file system APIs.

In this paper we discuss these interfaces and the options they provide, exercise DAOS through them with various I/O benchmarks, and analyse the observed performance. We also briefly compare the performance with a distributed file system and another object storage system deployed on the same hardware, and showcase DAOS' potential and increased flexibility to support high-performance I/O.

\end{abstract}


\section{Introduction}

The Distributed Asynchronous Object Store (DAOS)\cite{daos-scfa2022} is a novel software-defined object store designed for high-performance Input-Output (I/O) to distributed Non-Volatile Memory (NVM) devices. It currently provides a relevant alternative to traditional POSIX distributed/parallel file systems in an HPC context after having scored top positions in recent editions of the I/O 500\cite{io500-sc23}.

Distributed file systems need to be used carefully by applications to achieve optimal storage performance and ensure consistency of stored data under contention between reading and writing processes. Even then, there is a concern that these systems may not scale to extreme workloads due to fundamental design aspects, including metadata prescriptiveness, sustained involvement of the operating system, block-level operation, distributed locking to ensure client cache coherency, and potentially centralised metadata serving\cite{io-contention-filesystems}\cite{lockwood}\cite{lustre-internals}\cite{gpfs-internals}.

DAOS was designed from the ground up with these concerns in mind, and addresses them by having low metadata requirements on objects, operating fully in user space, providing byte-addressable access to data, implementing lockless contention resolution, and serving metadata in a fully distributed manner across all storage server nodes.

There are different interfaces provided for applications to interact with a DAOS system.

One of them is the libdaos library which provides an object storage API. This includes Array functionality, intended for bulk storage of large one-dimensio\-nal data arrays, and dictionary-like Key-Value functionality. Key-Values provide a mapping between keys (limited-length strings) and values (arbitrary-length data) that can be queried. Key-Value and Array objects must be created within a DAOS container, which provides an isolated object namespace and transaction history. Upon creation, objects are assigned a 128-bit unique object identifier (OID), of which 96 bits are user-managed. The sharding, replication, and erasure-coding of these objects can be controlled by specifying their object class on creation. If configured with any of these options, they are transparently stored entirely or by parts across different storage devices and nodes thus enabling efficient concurrent access.

Because porting existing POSIX I/O applications to libdaos is a daunting task, DAOS also provides the libdfs library which implements POSIX directories, files and symbolic links on top of the libdaos APIs. An application using this library can perform common file system calls which are transparently mapped to DAOS, with only minor modifications required. libdfs is not fully POSIX-compliant but supports the majority of existing POSIX-based applications.

A DAOS FUSE (file system in user space) daemon (DFUSE) is also provided which allows users to mount and expose a DAOS system through the standard POSIX infrastructure, enabling existing POSIX applications to operate on DAOS without modification. The DFUSE mount command supports several options, e.g. to specify the number of FUSE and event queue threads\cite{dfuse}, or to configure caching of file system data and metadata in the client nodes.

DFUSE can show limited performance under intensive small I/O workloads due to many round-trips required between kernel and user space. For these cases, an I/O interception library (IL), also provided with DAOS, can be used to forward operations directly to libdfs and reach optimal performance.

In this paper we present results from DAOS deployed on Google Cloud Platform (GCP) Virtual Machines (VMs; also referred to as instances or nodes) with locally attached NVMe SSDs, exercising the DAOS system through the available interfaces using a number of benchmark applications. We also tested the data protection features DAOS supports. With the obtained measurements, we analysed the performance and scalability of the different interfaces and options.

\newpage

\section{Methodology}

To analyse the performance DAOS can provide when accessed via the different available interfaces, we deployed DAOS on GCP instances with locally attached NVMe SSDs and ran a selection of benchmark applications, introduced in \ref{sec:benchmarks}, on separate instances. When run, the benchmarks output timestamp information that we used to calculate bandwidths and compare and discuss performance.

All selected benchmarks first execute a write phase followed by a read phase, and can be configured to run across multiple processes and nodes. Some benchmarks (e.g.\ IOR) can be configured such that each parallel process performs bulk I/O to a single object or file. As this I/O pattern executes small amounts of metadata operations and generally requires little logic to be triggered in the storage system, such benchmarks typically reach the maximum I/O bandwidths the storage system --- in this case DAOS --- can provide. We ran these benchmarks with DAOS data protection disabled, measured the achieved bandwidth, and compared this to the raw bandwidths available from the underlying storage and network hardware (as measured with lower-level tools).

Some of the benchmarks or configurations can trigger more complex I/O workloads with smaller object sizes or performing more metadata or Key-Value operations, and we used these to gain insight into the strengths and weaknesses of the storage system when under more realistic or complex workloads.

There are many options and parameters which may be adjusted when deploying DAOS on the server VMs and running the benchmarks on the client VMs, for which we made use of support from DAOS and Google Cloud experts. For the exploration of parameters in the benchmark runs, we tested every benchmark with different client node and process counts to determine the maximum achievable bandwidth, as outlined in \cite{daos-ipdps} and \cite{daos-lustre-pasc}. We then ran all benchmarks using the optimal node and process counts against DAOS servers deployed on increasing numbers of instances, to assess the scalability of DAOS and the benchmarks. Finally, we reran some of the benchmarks with DAOS data redundancy enabled.

Each benchmark run took between 1 and 20 minutes of wall-clock time, depending on the scale, thus providing insight on short-term performance rather than long-term stability.

We applied a common bandwidth definition for the calculation of all bandwidths reported in \ref{sec:results}: the amount of data transferred (written or read) divided by the wall-clock time elapsed between the start of the first I/O operation and the end of the last I/O operation. Each and every test was repeated 3 times, and the average and standard deviation of the measured bandwidths are shown in the figures for each test.

\subsection{Benchmarks}
\label{sec:benchmarks}

\subsubsection{IOR\cite{ior}}
is a popular open-source I/O benchmark developed by the HPC community, originally intended to measure the I/O performance of parallel file systems, but expanded over time with new I/O backends to support operation on other storage systems like DAOS and Ceph. It runs as a parallel MPI application, where the concurrent processes create a file or object each, wait for each other, and commence issuing a sequence of write or read operations. IOR can be configured to have all processes operate against a single shared file, a file per process, to adjust the number and size of operations per process and their distribution in the file, and to reproduce other common I/O patterns and approaches.

\subsubsection{HDF5\cite{hdf5}}
is a library for efficient storage of complex and voluminous data sets used in a range of disciplines, and supports features such as data compression and encryption. HDF5 operates, by default, on POSIX file systems, but has adaptors to support operation on other storage systems including DAOS\cite{hdf5-daos-vol}. The IOR benchmark has a backend that uses HDF5, allowing benchmarking of HDF5 approaches with IOR. When IOR is run with the HDF5 backend on POSIX, a file is created per writer process, where the process metadata, indexing information, and data are stored. If the DAOS adaptor is enabled, a DAOS container is created per writer process, and the data from every write operation stored in a separate object in the container. This makes the HDF5 approach with IOR a more complex set of I/O patterns than those performed when using POSIX or DAOS backends.

\subsubsection{Field I/O\cite{field-io}}
is a standalone benchmark tool developed by ECMWF to evaluate the performance a DAOS system can provide for Numerical Weather Prediction (NWP) operations at the centre, without involving the full complexity of their operational I/O stack. It runs as a set of independent processes, each writing and indexing a sequence of weather variables, or fields, into DAOS with a combination of libdaos Array and Key-Value operations. If configured in read mode, the processes retrieve the same sequence of fields by querying the Key-Values and reading the Array data. Field I/O processes write each field in a separate Array, and store indexing information in a set of Key-Values --- some of them exclusive to the process, and some of them shared amongst all processes.

\subsubsection{fdb-hammer\cite{fdb-hammer}}
is a benchmark tool to measure the performance of ECMWF's domain-specific object store, the FDB\cite{fdb-pasc2019}. FDB implements transactional and efficient weather field storage and indexing on a number of storage systems, including POSIX file systems, DAOS, and Ceph. FDB exposes a scientifically meaningful API for applications to archive and retrieve weather fields without requiring knowledge of the underlying storage system, effectively abstracting it away.

fdb-hammer runs as a set of independent processes, each archiving or retrieving (depending on the selected access mode) a sequence of weather fields via FDB. When run with FDB's DAOS backend, fdb-hammer uses a set of libdaos Arrays and Key-Values to store and index the weather fields, in a similar way as Field I/O. When run with FDB's POSIX backend, fdb-hammer writer processes create a pair of files each, which are expanded incrementally with indexing information and field data, respectively. Writer processes accumulate small chunks of data in client memory, that are persisted periodically into the file system in large blocks to achieve optimal write performance, with the aim of avoiding throttling ECMWF's NWP model in operations. Reader processes repeatedly open and read, for every field in the sequence, the corresponding files containing the index and field data, resulting in substantial metadata and small I/O operation workloads.

\subsection{Test system}
\label{sec:testsystem}

All tests were conducted on Google Cloud Platform (GCP) infrastructure. For the DAOS deployments we used VMs of a custom type\cite{gcp-instance-types} n2-custom-36-153600, each with 36 logical cores, 150 GiB of DRAM, 6 TiB of local NVMe SSDs dis\-tributed in 16 logical devices, and a 50 Gbps network adaptor.

The benchmarks were run in VMs of type n2-highcpu-32, each with 32 logical cores, 32 GiB of DRAM, and a 50 Gbps network adaptor. To exploit all available network bandwidth, we carefully pinned the benchmark processes evenly across all available cores, and configured all VMs with Simultaneous Multi-Threading (SMT)\cite{gcp-smt} enabled.

DAOS v2.4.1 libraries were installed both on the server and client side. The DAOS deployments had a single engine deployed per server VM, and each engine was configured to deploy 16 DAOS targets\cite{daos-storage-model} in the VM --- one per available NVMe SSD. DAOS was configured to use the DRAM in the VMs for metadata storage as there was no Storage Class Memory in these VMs.

Although GCP provides generic VMs, not necessarily hard\-ware-tailored purely for HPC, it makes an excellent standard testbed to analyse software-side performance and scalability of storage systems such as DAOS. The same overall behaviour should be expected on systems using similar technology.

\section{Results}
\label{sec:results}

\subsection{Hardware Bandwidth}

The raw bandwidth of the NVMe SSDs on server instances for bulk I/O was measured by mounting each of the 16 drives in one of the instances as an ext4 file system and then running the \texttt{dd} command in parallel for all of them, first writing and then reading 1000 blocks of 100 MiB. The measurements showed 3.86 GiB/s of aggregate write bandwidth and 7 GiB/s of aggregate read bandwidth.

\texttt{iperf} was used to measure raw network bandwidth between client and server instances, which was found to match the expected 50 Gbps (6.25 GiB/s) in both directions.

These measurements indicated that every additional DAOS server instance employed in a DAOS deployment could at best provide an additional 3.86 GiB/s for write, limited by the SSD bandwidth, and 6.25 GiB/s for read, limited by the network.

\subsection{API performance comparison}

\begin{figure}[htbp]
    \centering
    \begin{subfigure}[b]{126pt}
        \includegraphics[width=126pt,trim={0 16pt 0 0},clip]{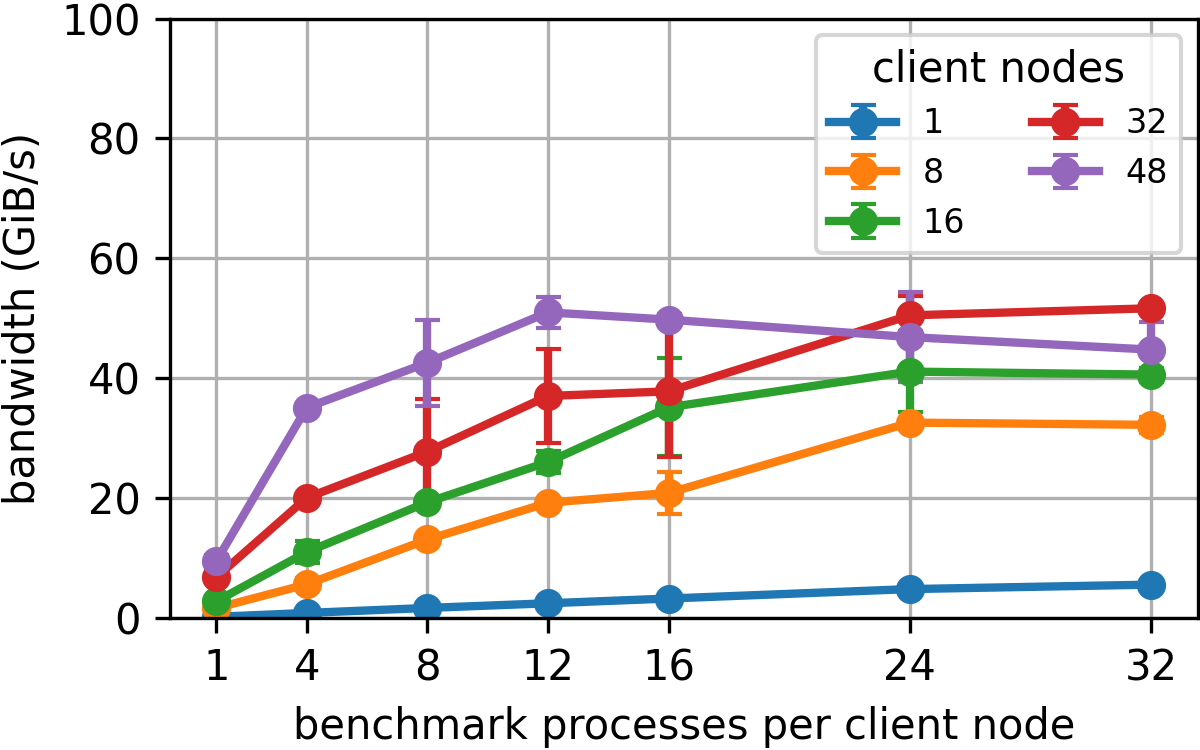}
        \caption{libdaos, Write}
    \end{subfigure}
    \begin{subfigure}[b]{110pt}
        \includegraphics[width=110pt,trim={35pt 16pt 0 0},clip]{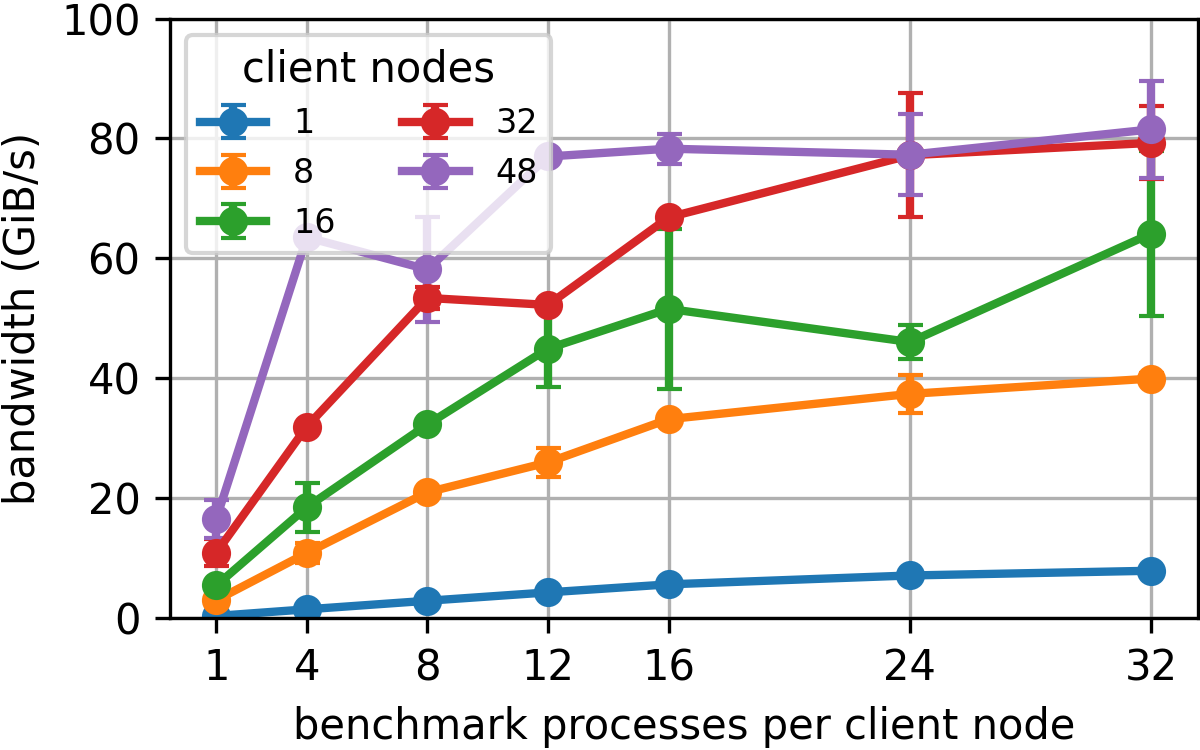}
        \caption{libdaos, Read}
    \end{subfigure}
    \begin{subfigure}[b]{126pt}
        \vspace{6pt}
        \includegraphics[width=126pt,trim={0 16pt 0 0},clip]{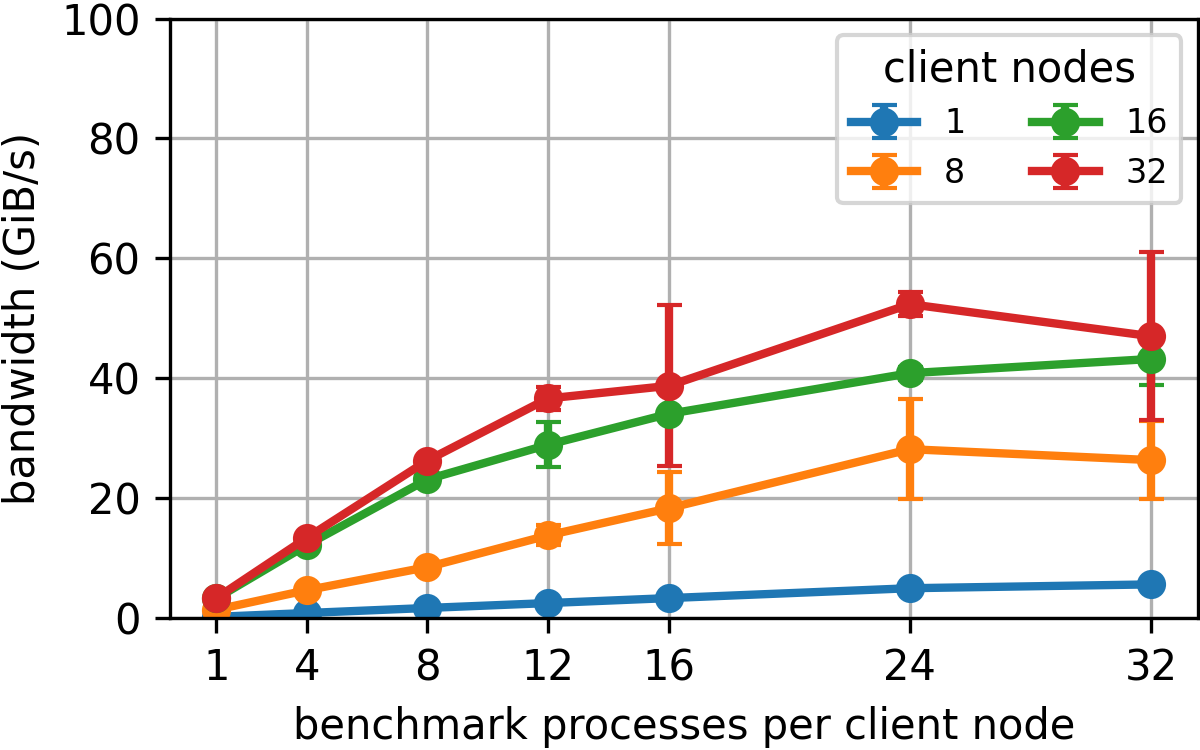}
        \caption{libdfs, Write}
    \end{subfigure}
    \begin{subfigure}[b]{110pt}
        \vspace{6pt}
        \includegraphics[width=110pt,trim={35pt 16pt 0 0},clip]{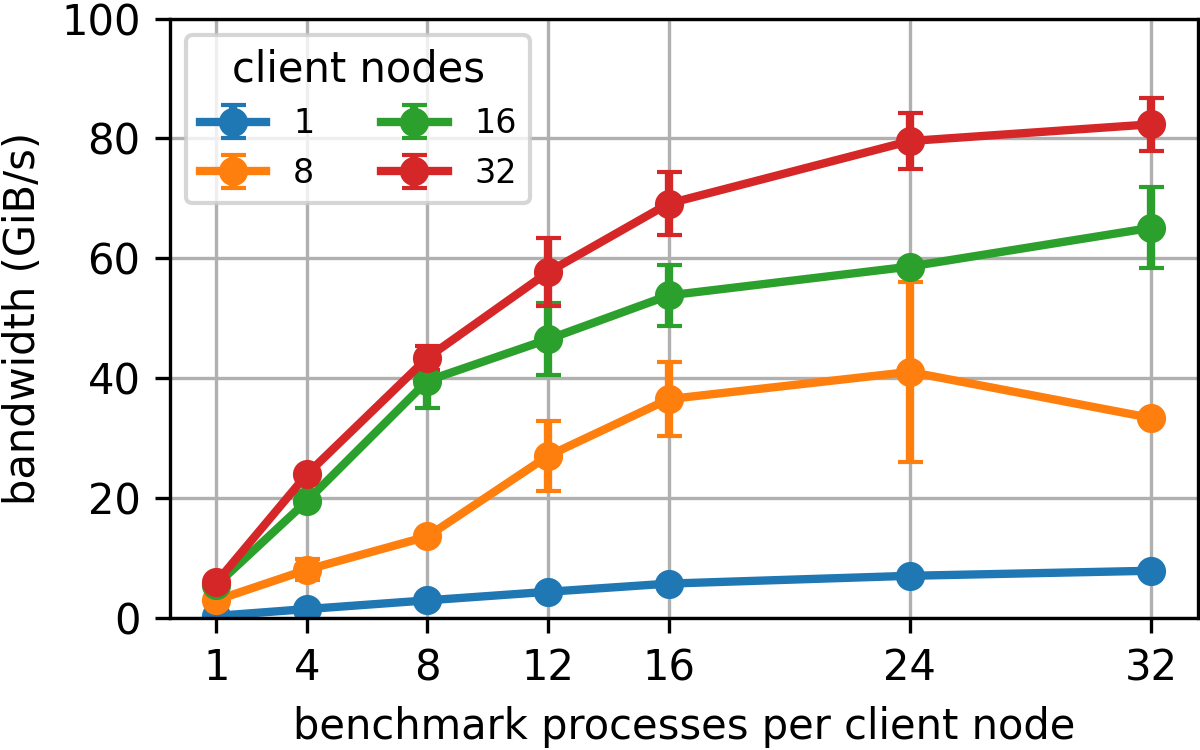}
        \caption{libdfs, Read}
    \end{subfigure}
    \begin{subfigure}[b]{126pt}
        \vspace{6pt}
        \includegraphics[width=126pt,trim={0 16pt 0 0},clip]{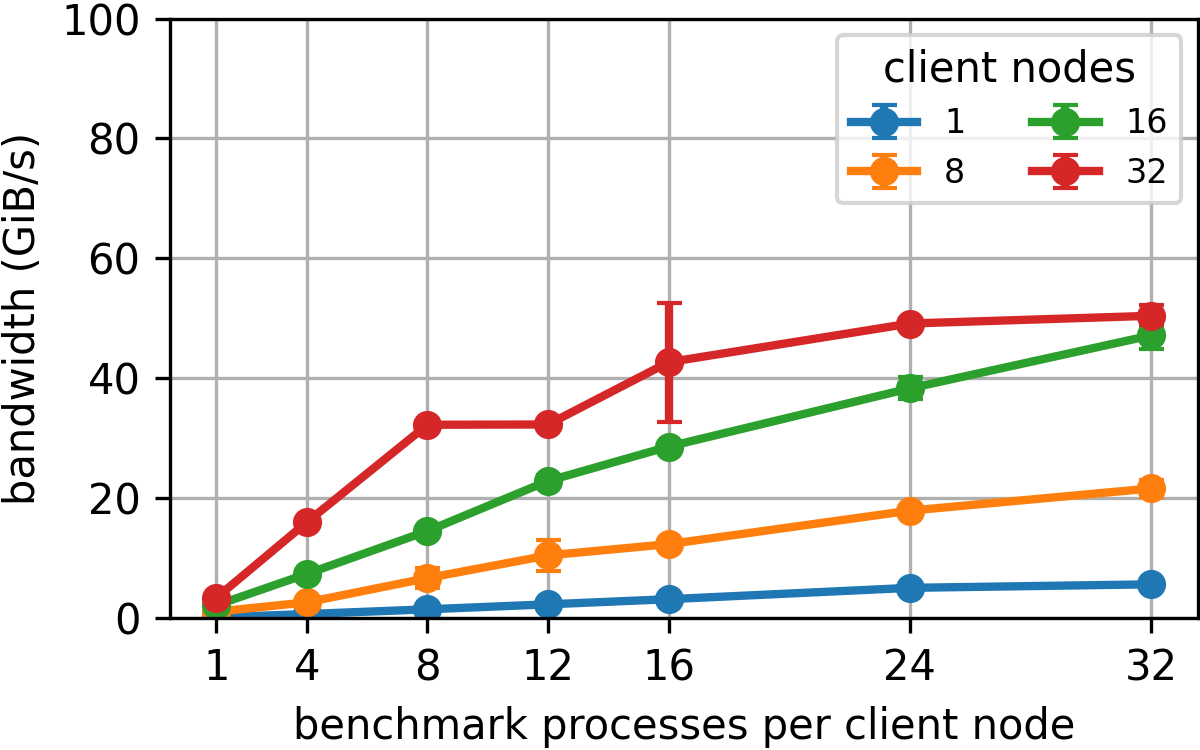}
        \caption{DFUSE, Write}
    \end{subfigure}
    \begin{subfigure}[b]{110pt}
        \vspace{6pt}
        \includegraphics[width=110pt,trim={35pt 16pt 0 0},clip]{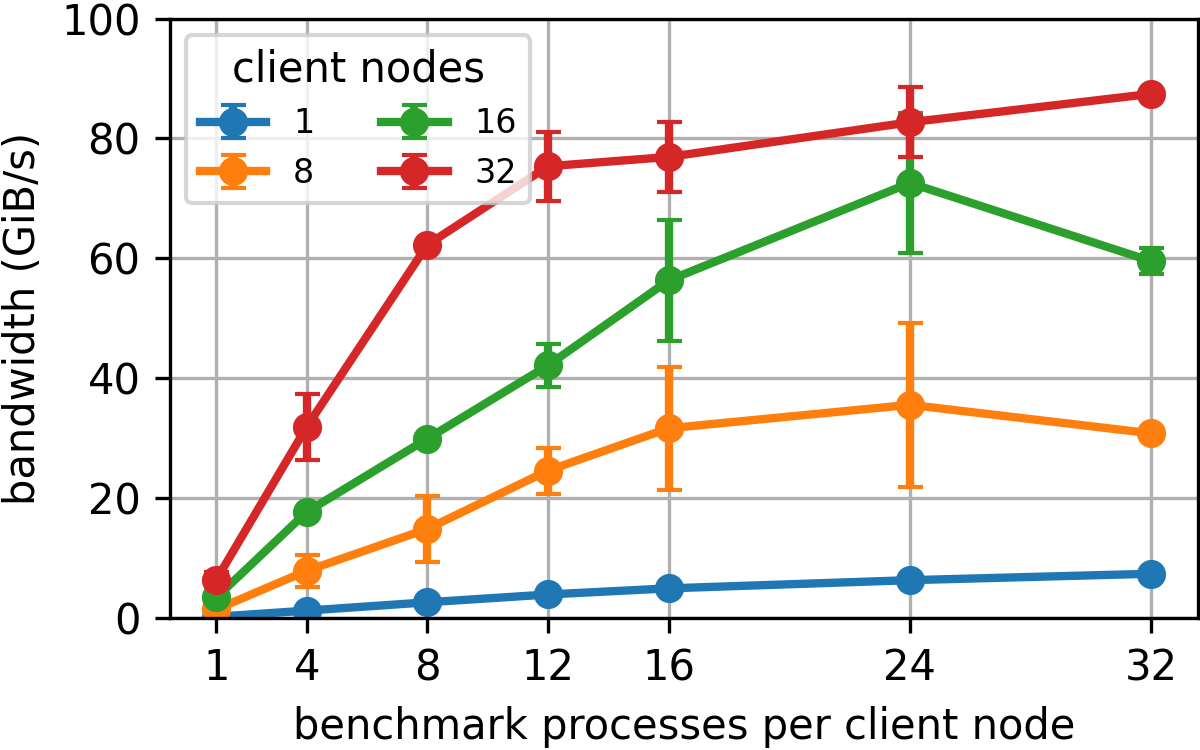}
        \caption{DFUSE, Read}
    \end{subfigure}
    \begin{subfigure}[b]{126pt}
        \vspace{6pt}
        \includegraphics[width=126pt]{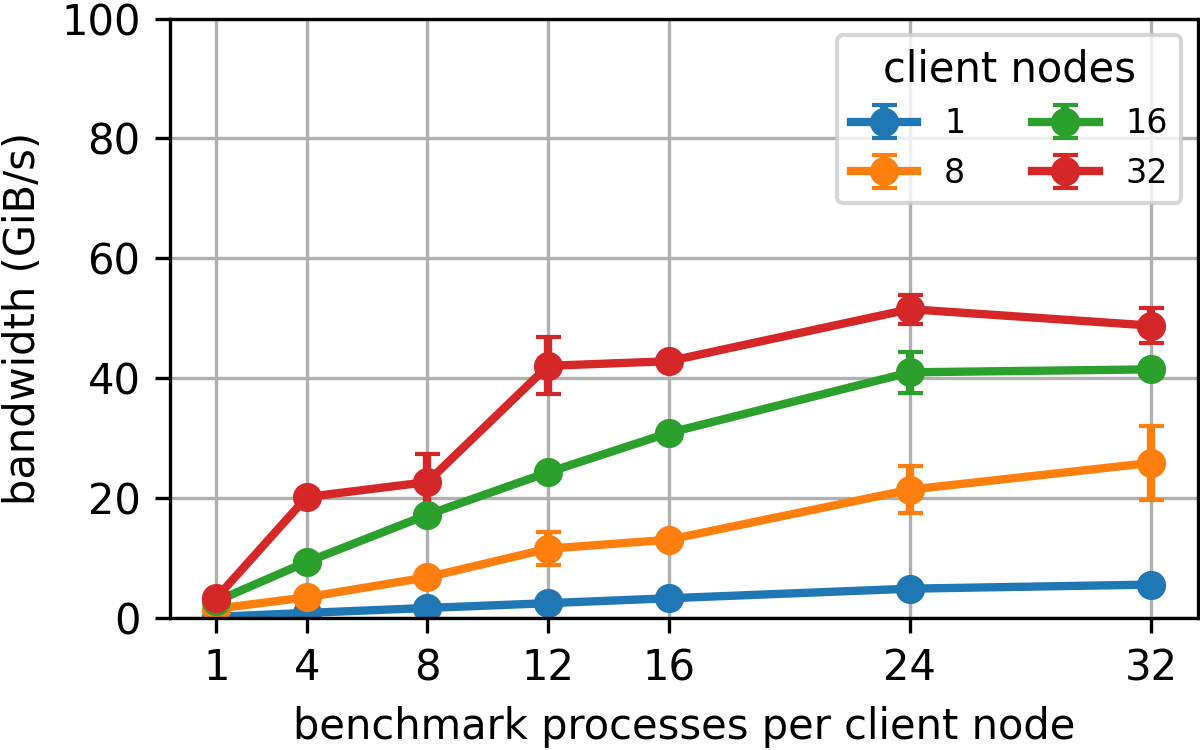}
        \caption{DFUSE+IL, Write}
    \end{subfigure}
    \begin{subfigure}[b]{110pt}
        \vspace{6pt}
        \includegraphics[width=110pt,trim={35pt 0 0 0},clip]{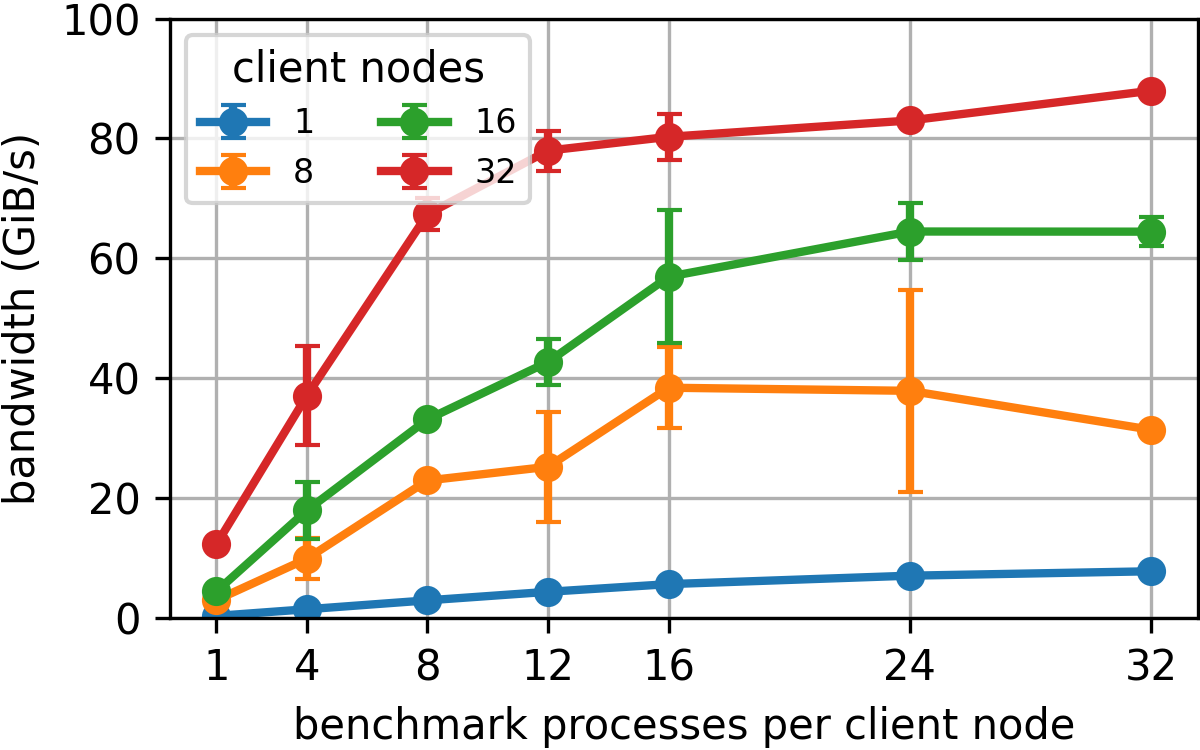}
        \caption{DFUSE+IL, Read}
    \end{subfigure}
    \caption{Client node and process count optimisation results for IOR with the different DAOS APIs, against a 16-node DAOS instance.}
    \label{fig:ior_16sn_cn_cpcn}
\end{figure}

We deployed a DAOS system on 16 nodes with NVMe and ran IOR with the libdaos and libdfs backends, and with the POSIX backend on DFUSE and DFUSE with interception, varying client node and process count. IOR was configured to have each parallel process run a sequence of 10000 I/O operations (write or read) each of 1 MiB in a single Array or file per process. We selected an object class of \texttt{SX} (sharding across all targets) for the libdaos Arrays and files and directories in libdfs and DFUSE, as this was found to perform best. For DFUSE mounts, we configured 24 FUSE threads and 12 event queue threads, and disabled all caching. The results are shown in Fig. \ref{fig:ior_16sn_cn_cpcn}.

Although there are slight differences in the performance behaviour with respect to the process count --- i.e. the libdaos backend achieves higher bandwidths at lower process counts --- all APIs generally behave similarly and can reach bandwidths of nearly 60 GiB/s for write and 90 GiB/s for read. These bandwidths are very close to the calculated optimum of 61.76 GiB/s for write and 112 GiB/s for read. 16 client nodes, that is, a ratio of 1-to-1 client-to-server nodes, are generally sufficient to approach the best bandwidths.

It is worth noting that the DFUSE interception library does not bring much benefit with the selected I/O size of 1 MiB. The benefit becomes very noticeable for smaller I/O sizes, as shown in Fig. \ref{fig:ior_dfuse_dfuseil_16sn_cn_cpcn_small}, comparing the amount of I/O operations per second (IOPS) reached with DFUSE and DFUSE with IL using an I/O size of 1 KiB.

\begin{figure}[htbp]
    \centering
    \begin{subfigure}[b]{124pt}
        \includegraphics[width=124pt,trim={0 16pt 0 0},clip]{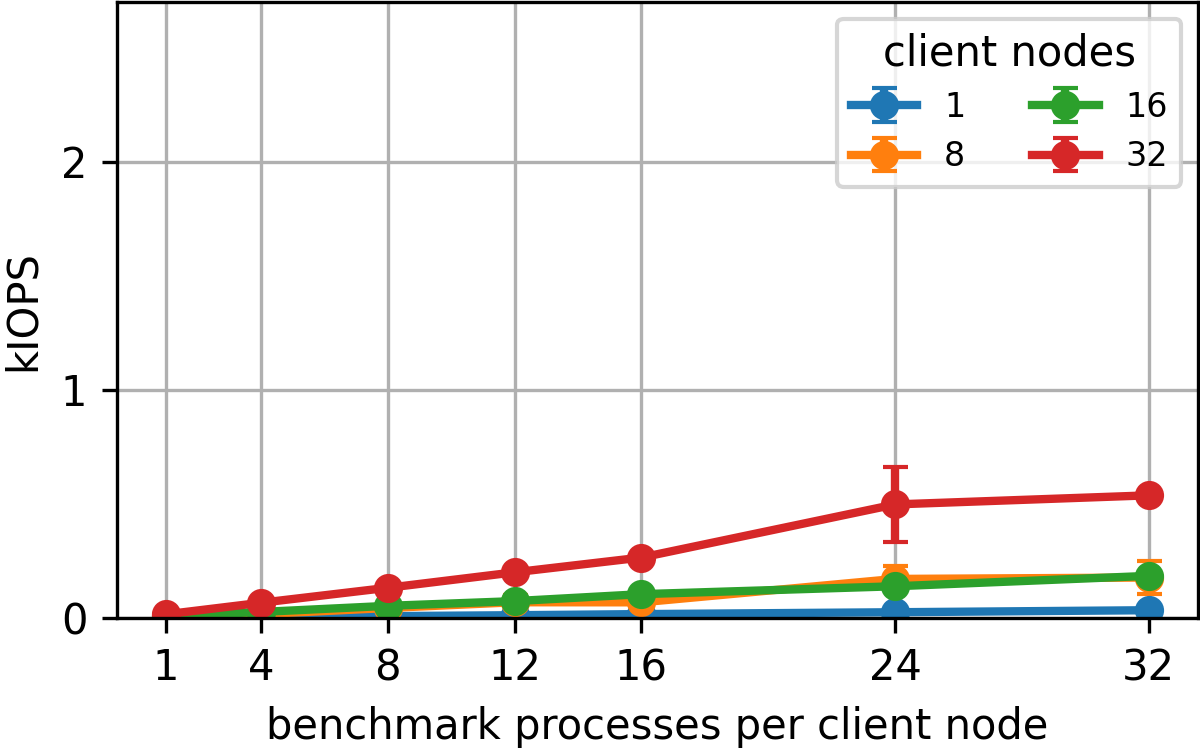}
        \caption{DFUSE, Write}
    \end{subfigure}
    \begin{subfigure}[b]{115pt}
        \includegraphics[width=115pt,trim={21pt 16pt 0 0},clip]{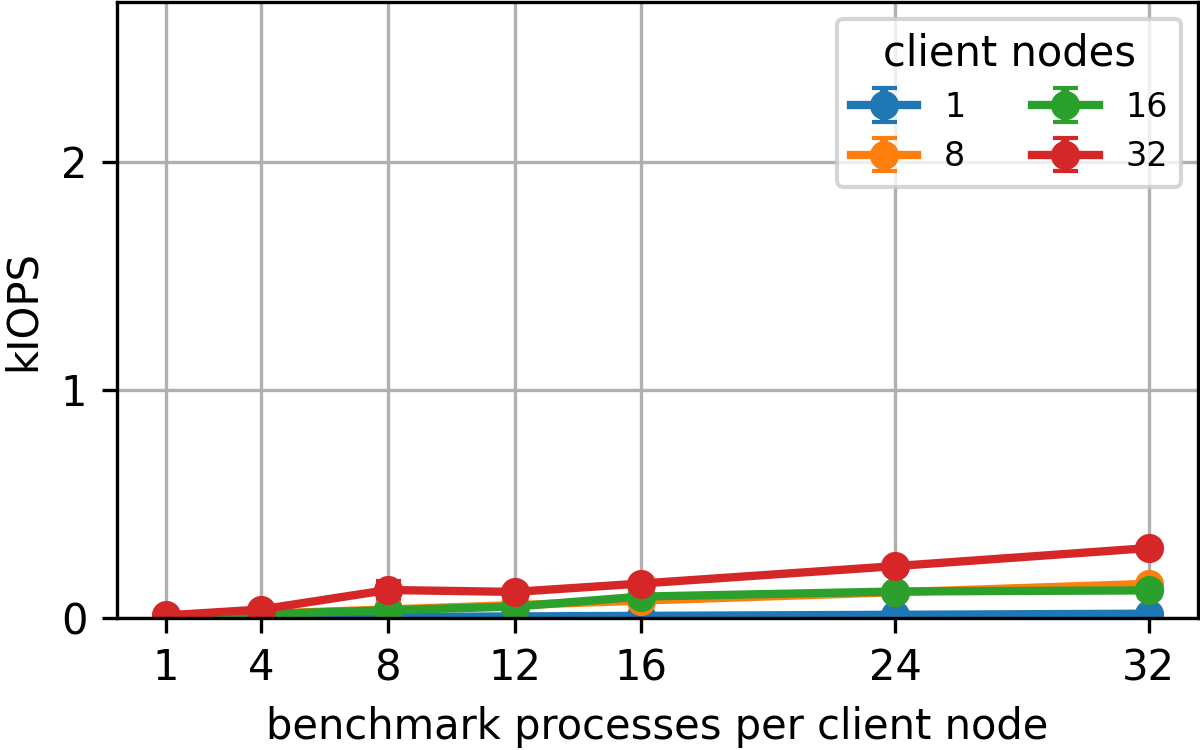}
        \caption{DFUSE, Read}
    \end{subfigure}
    \begin{subfigure}[b]{124pt}
        \vspace{6pt}
        \includegraphics[width=124pt]{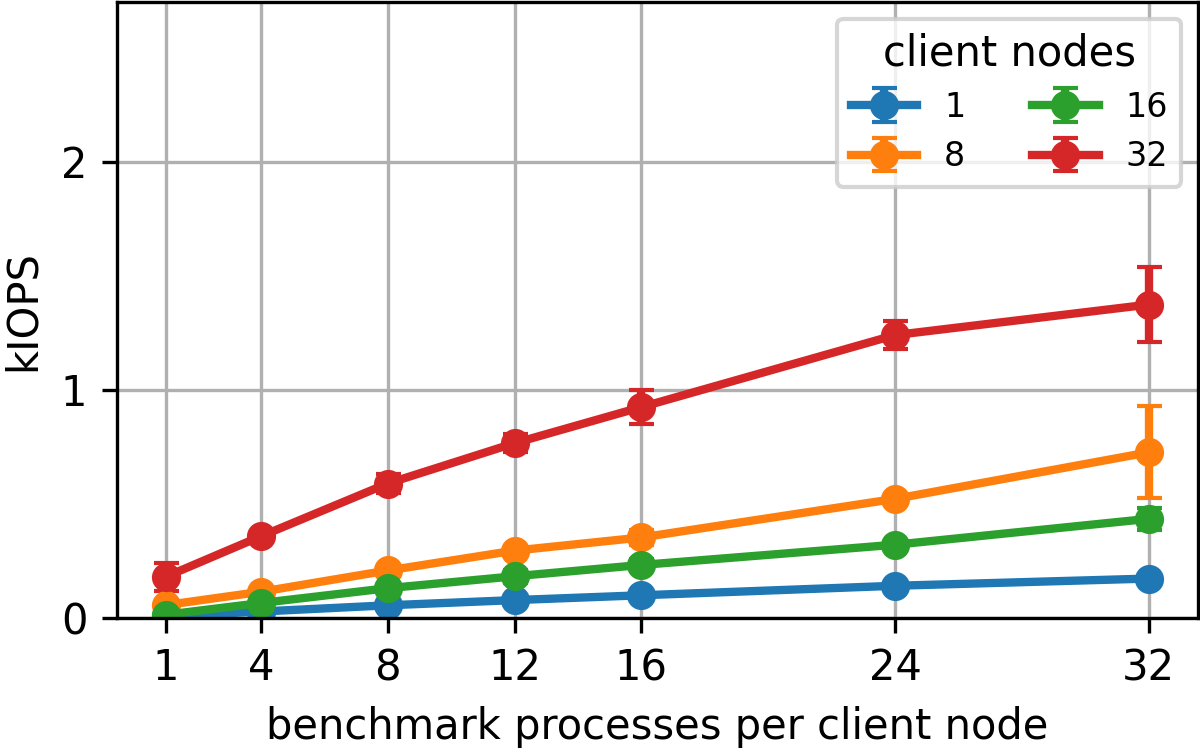}
        \caption{DFUSE+IL, Write}
    \end{subfigure}
    \begin{subfigure}[b]{115pt}
        \vspace{6pt}
        \includegraphics[width=115pt,trim={21pt 0 0 0},clip]{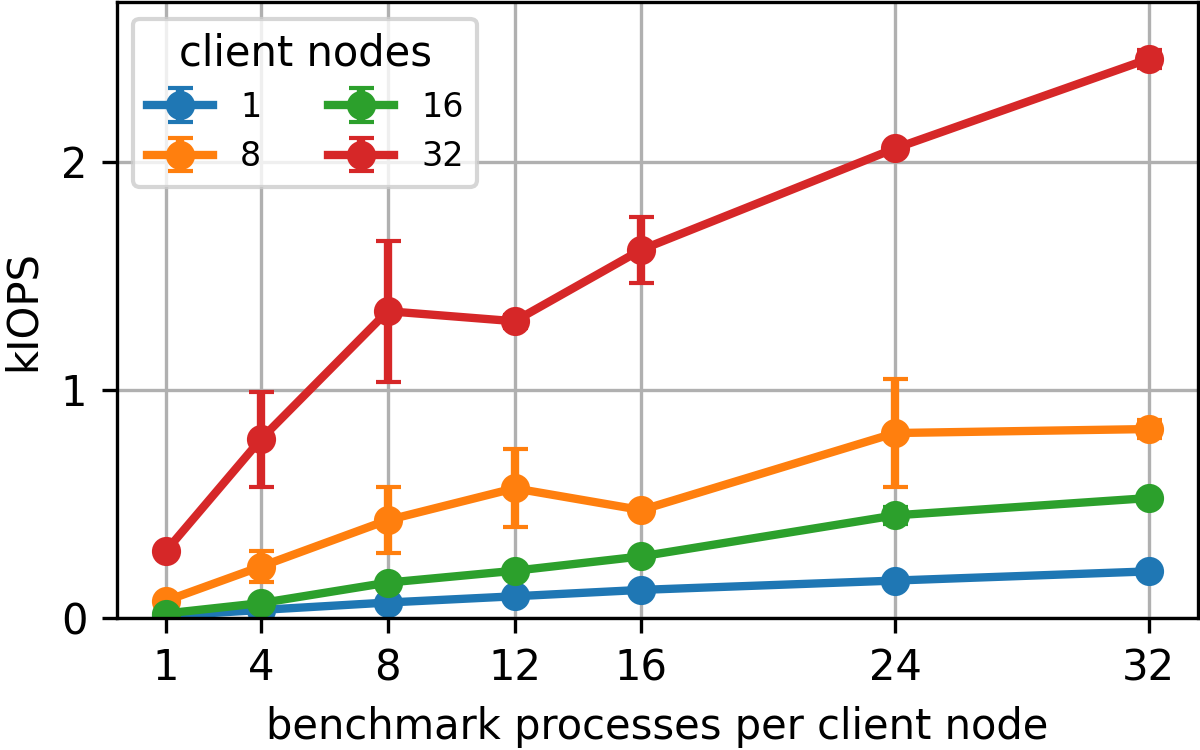}
        \caption{DFUSE+IL, Read}
    \end{subfigure}
    \caption{Client node and process count optimisation results for IOR on DFUSE and DFUSE+IL with 1KiB I/O size, against a 16-node DAOS instance.}
    \label{fig:ior_dfuse_dfuseil_16sn_cn_cpcn_small}
\end{figure}

We then ran similar tests using the more complex benchmarks, including IOR with the HDF5 backend using POSIX through DAOS DFUSE with interception, IOR with the HDF5 backend configured with the DAOS adaptor operating natively on libdaos, and Field I/O and fdb-hammer both operating natively using libdaos. All were configured to perform equivalent I/O workloads as in previous IOR tests, that is, performing 10k I/O operations of 1 MiB per process. While HDF5 on POSIX uses a file per process, the rest of the applications use a separate DAOS Array for every I/O. We found the optimum object classes to be \texttt{SX} for files in HDF5 on POSIX and objects in HDF5 on libdaos; \texttt{SX} for Key-Values and \texttt{S1} for Arrays in Field I/O; and \texttt{S1} both for Key-Values and Arrays in fdb-hammer. The results are shown in Fig. \ref{fig:apps_16sn_cn_cpcn}.

\begin{figure}[htbp]
    \centering
    \begin{subfigure}[b]{126pt}
        \includegraphics[width=126pt,trim={0 16pt 0 0},clip]{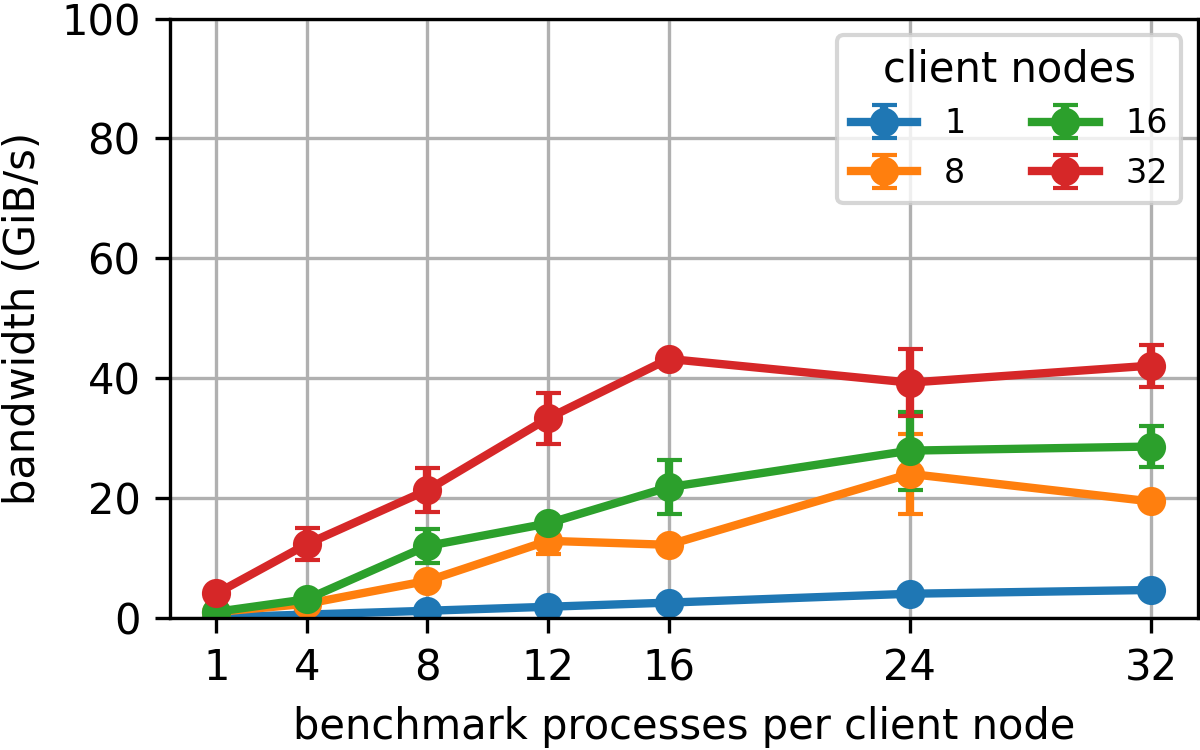}
        \caption{IOR/HDF5 on DFUSE+IL, Write}
    \end{subfigure}
    \begin{subfigure}[b]{110pt}
        \includegraphics[width=110pt,trim={35pt 16pt 0 0},clip]{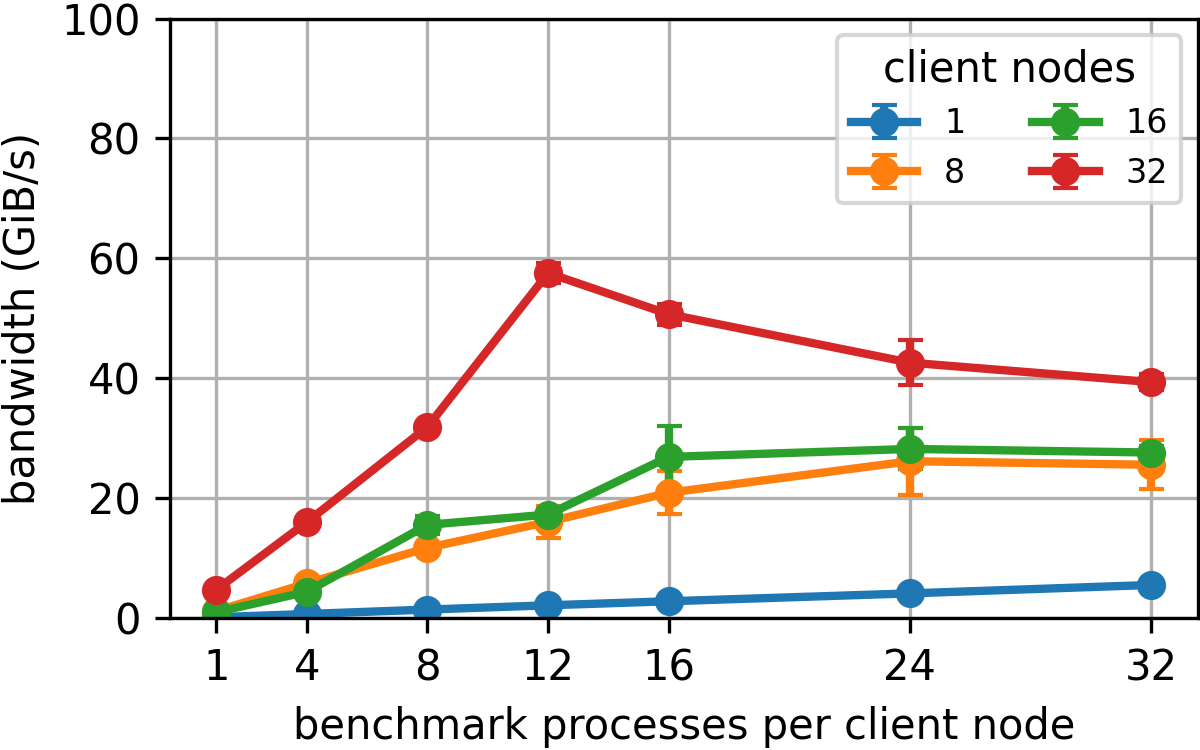}
        \caption{IOR/HDF5 DFUSE+IL, Read}
    \end{subfigure}
    \begin{subfigure}[b]{126pt}
        \vspace{6pt}
        \includegraphics[width=126pt,trim={0 16pt 0 0},clip]{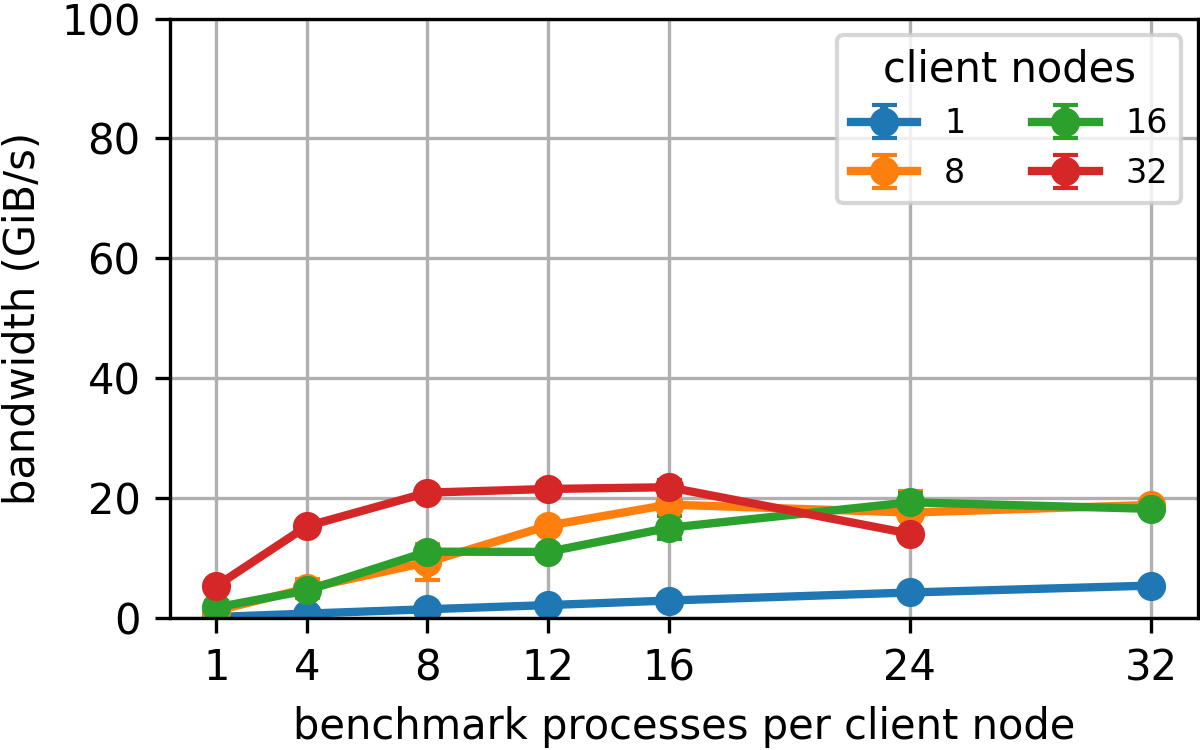}
        \caption{IOR/HDF5 on libdaos, Write}
    \end{subfigure}
    \begin{subfigure}[b]{110pt}
        \vspace{6pt}
        \includegraphics[width=110pt,trim={35pt 16pt 0 0},clip]{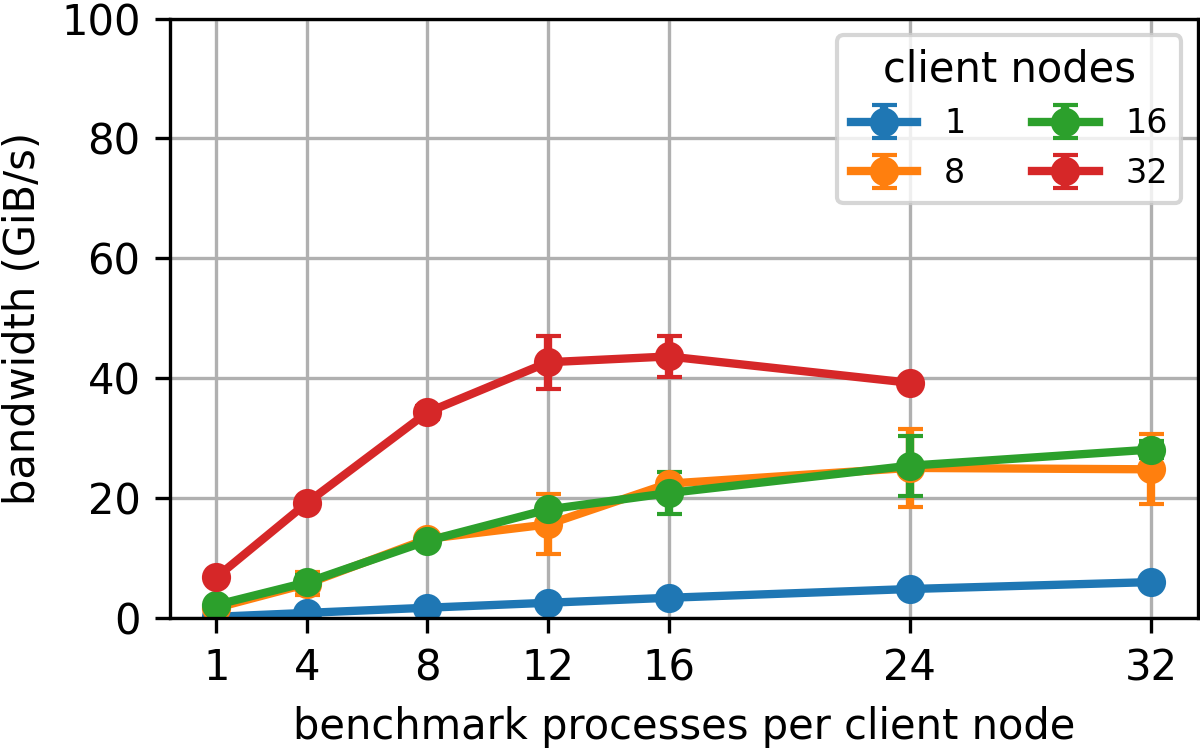}
        \caption{IOR/HDF5 on libdaos, Read}
    \end{subfigure}
    \begin{subfigure}[b]{126pt}
        \vspace{6pt}
        \includegraphics[width=126pt,trim={0 16pt 0 0},clip]{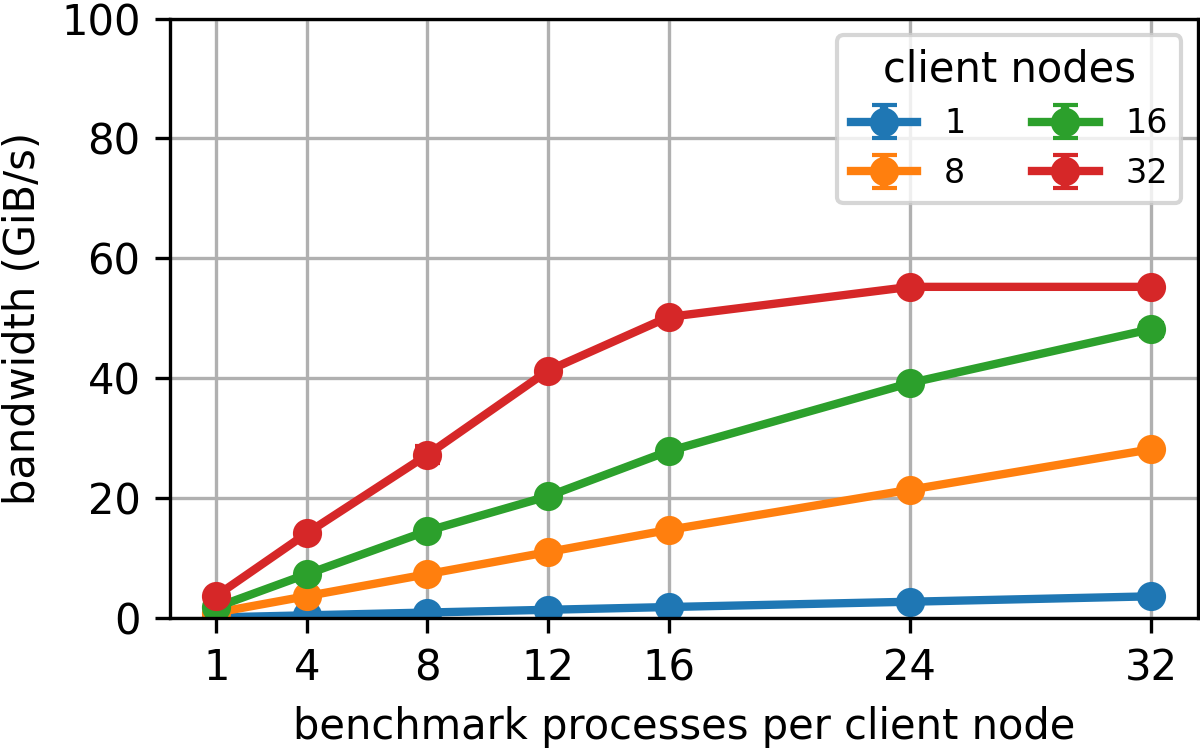}
        \caption{Field I/O on libdaos, Write}
    \end{subfigure}
    \begin{subfigure}[b]{110pt}
        \vspace{6pt}
        \includegraphics[width=110pt,trim={35pt 16pt 0 0},clip]{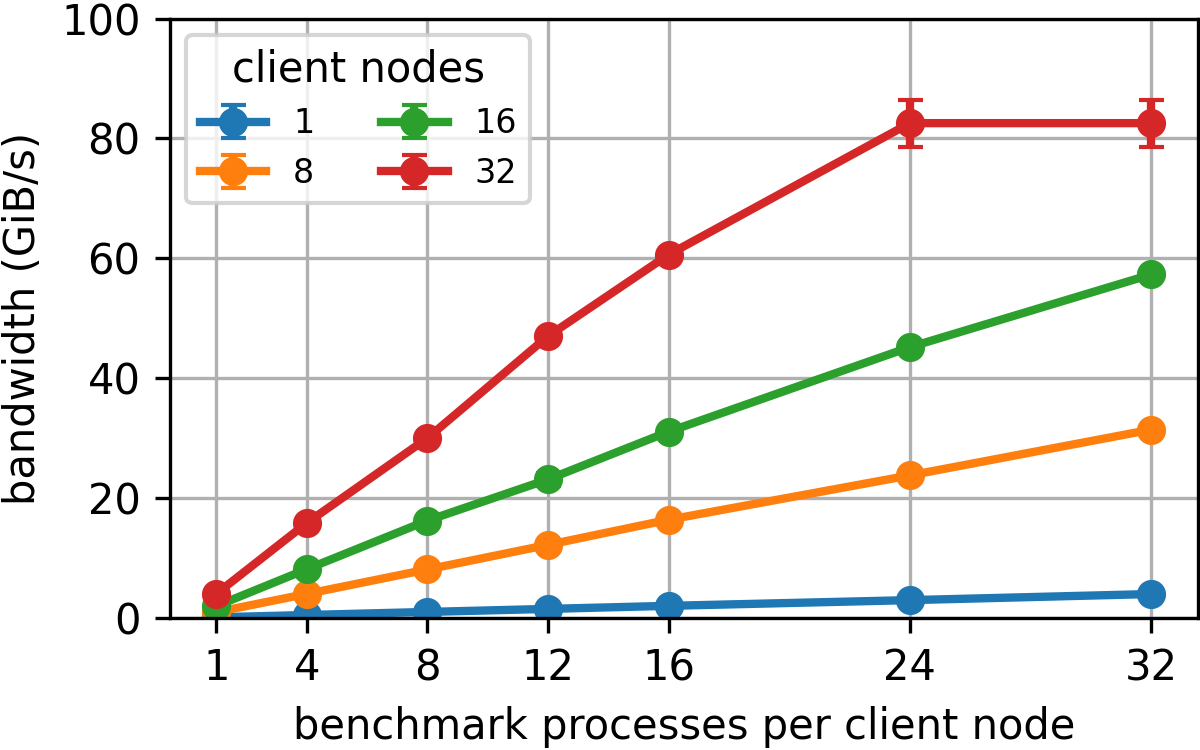}
        \caption{Field I/O on libdaos, Read}
    \end{subfigure}
    \begin{subfigure}[b]{126pt}
        \vspace{6pt}
        \includegraphics[width=126pt]{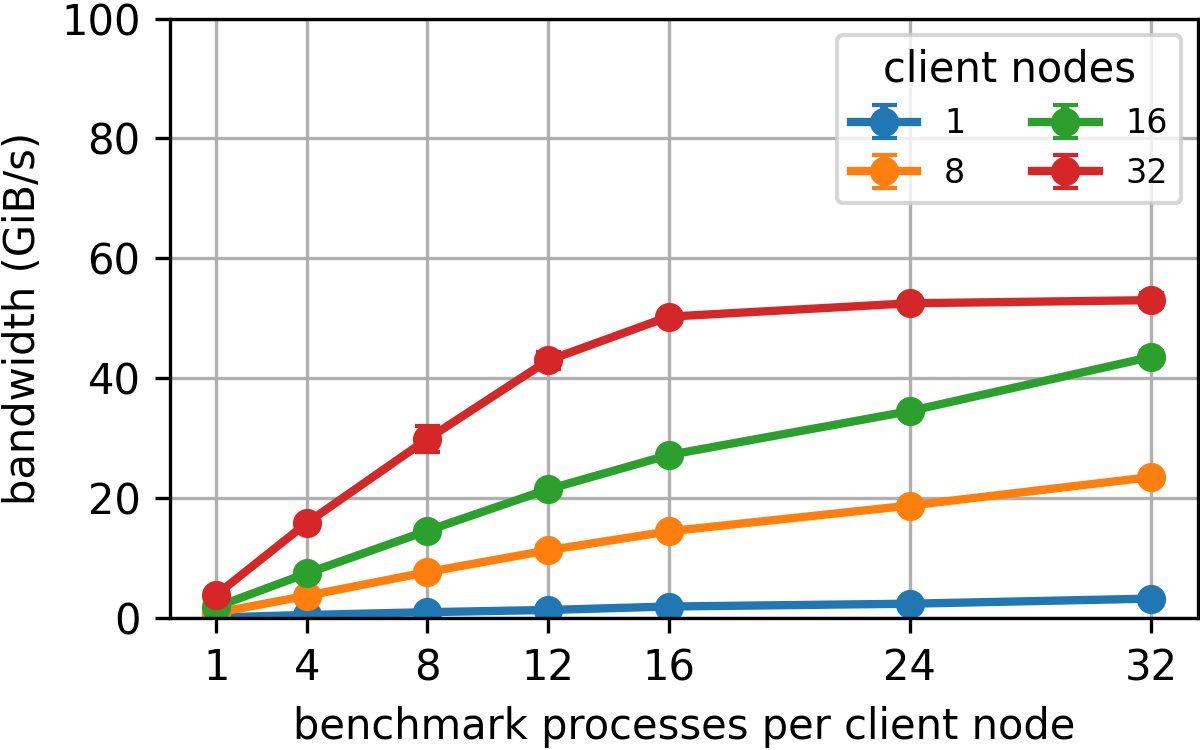}
        \caption{fdb-hammer on libdaos, Write}
    \end{subfigure}
    \begin{subfigure}[b]{110pt}
        \vspace{6pt}
        \includegraphics[width=110pt,trim={35pt 0 0 0},clip]{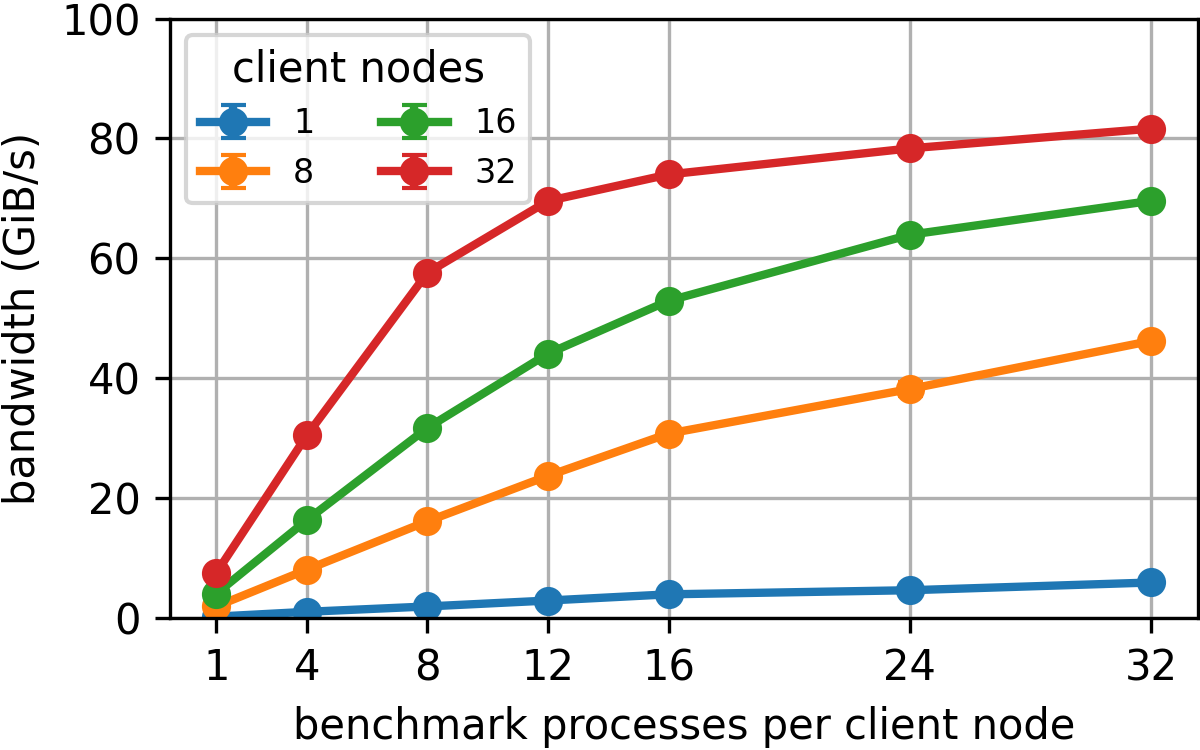}
        \caption{fdb-hammer on libdaos, Read}
    \end{subfigure}
    \caption{Client node and process count optimisation results for the different DAOS applications, against a 16-node DAOS instance.}
    \label{fig:apps_16sn_cn_cpcn}
\end{figure}

Fig. \ref{fig:apps_16sn_cn_cpcn} (e), (f), (g) and (h) show that both the Field I/O and fdb-hammer benchmarks perform very well, with similar I/O performance compared to the simple IOR runs. This is a great result showing that DAOS enables performance at scale even if using relatively small transfer and object sizes, and performing many Key-Value operations. The two benchmarks perform an average of 10 Key-Value operations (put or get) for each of the 10k objects accessed by each process, to provide a domain-appropriate index of the data written.

The bandwidth of the read mode in Field I/O increases linearly as the processes count is increased. This scaling is inferior to that shown by fdb-hammer. This is likely explained by the fact that fdb-hammer contains optimisations to avoid object size checks prior to every read operation, and Field I/O does not implement these.

HDF5 runs on DFUSE and libdaos, in Fig. \ref{fig:apps_16sn_cn_cpcn} (a), (b), (c) and (d), show inferior bandwidths than the rest of the benchmarks, particularly HDF5 on libdaos. Similar runs for HDF5 on libdaos against a 4-node DAOS system, in Fig. \ref{fig:ior_hdf5_libdaos_4sn_cn_cpcn} (c) and (d), show that HDF5 on libdaos can approach optimal hardware performance similarly to IOR on libdaos as shown in Fig. \ref{fig:ior_hdf5_libdaos_4sn_cn_cpcn} (a) and (b). This points to a potential scalability issue in the HDF5 DAOS adaptor, as it can perform well at small scale but not at large scales. The scalability issue is likely to lie in the fact that the adaptor uses a DAOS container per parallel writing process, and this harms performance at large scales in DAOS as demonstrated in \cite{daos-ipdps}.

\subsection{Scalability of DAOS and the APIs}

We ran all benchmarks using the optimal client node and process counts, as determined in the previous section, against DAOS deployments on increasing numbers of nodes. The results, in Fig. \ref{fig:scalability_no_redun}, show that DAOS --- when used via all APIs --- and most applications can scale approximately linearly up to 24 DAOS server nodes, reaching close to ideal performance.

\begin{figure}[htbp]
    \centering
    \begin{subfigure}[b]{124pt}
        \includegraphics[width=124pt,trim={0 16pt 0 0},clip]{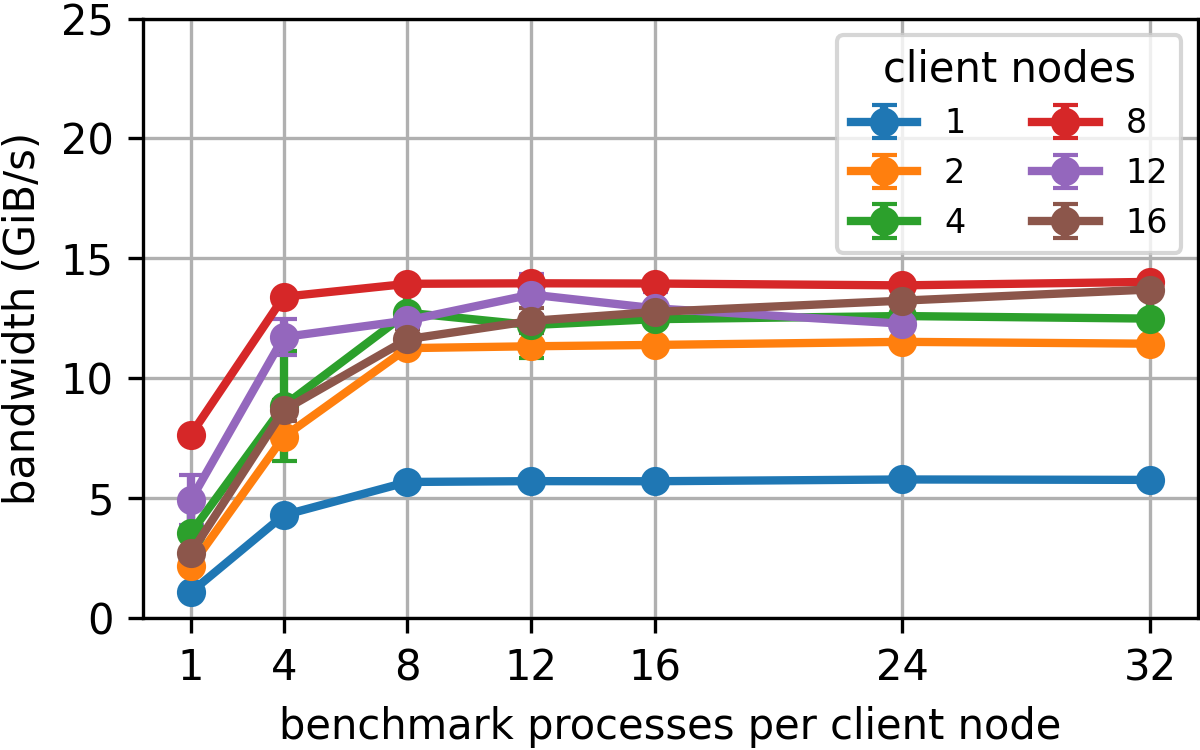}
        \caption{IOR on libdaos, Write}
    \end{subfigure}
    \begin{subfigure}[b]{112pt}
        \includegraphics[width=112pt,trim={27pt 16pt 0 0},clip]{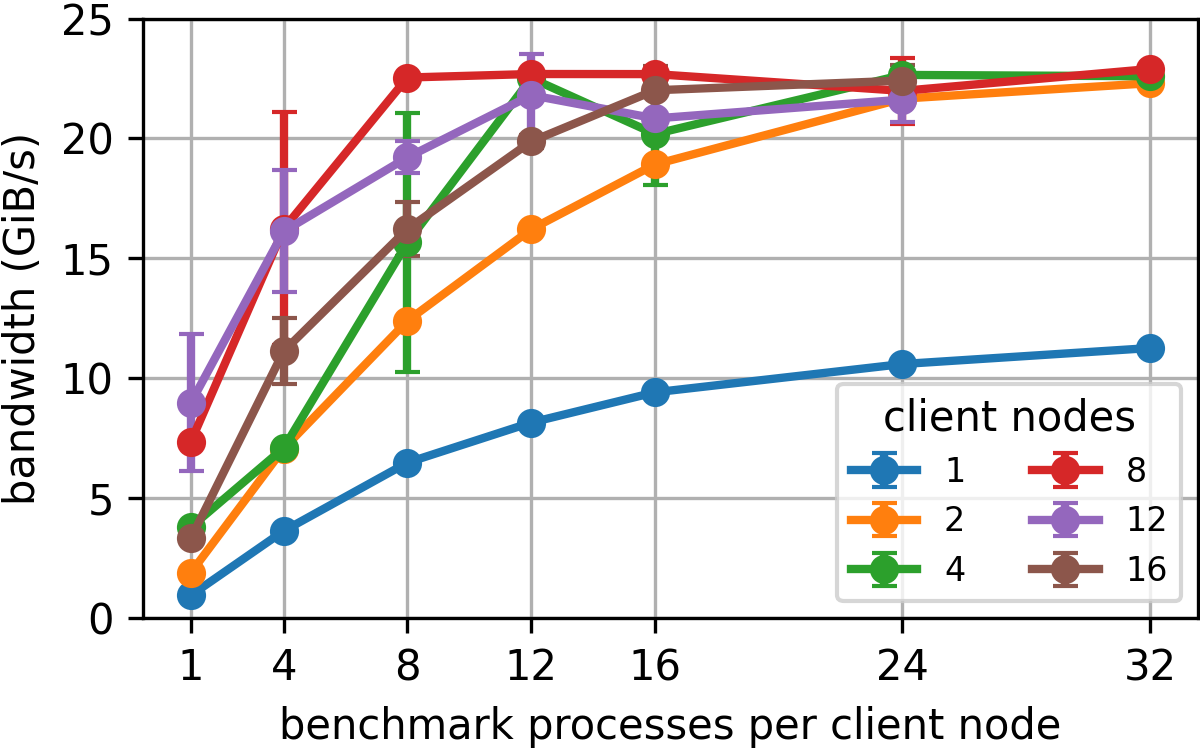}
        \caption{IOR on libdaos, Read}
    \end{subfigure}
    \begin{subfigure}[b]{124pt}
        \vspace{6pt}
        \includegraphics[width=124pt]{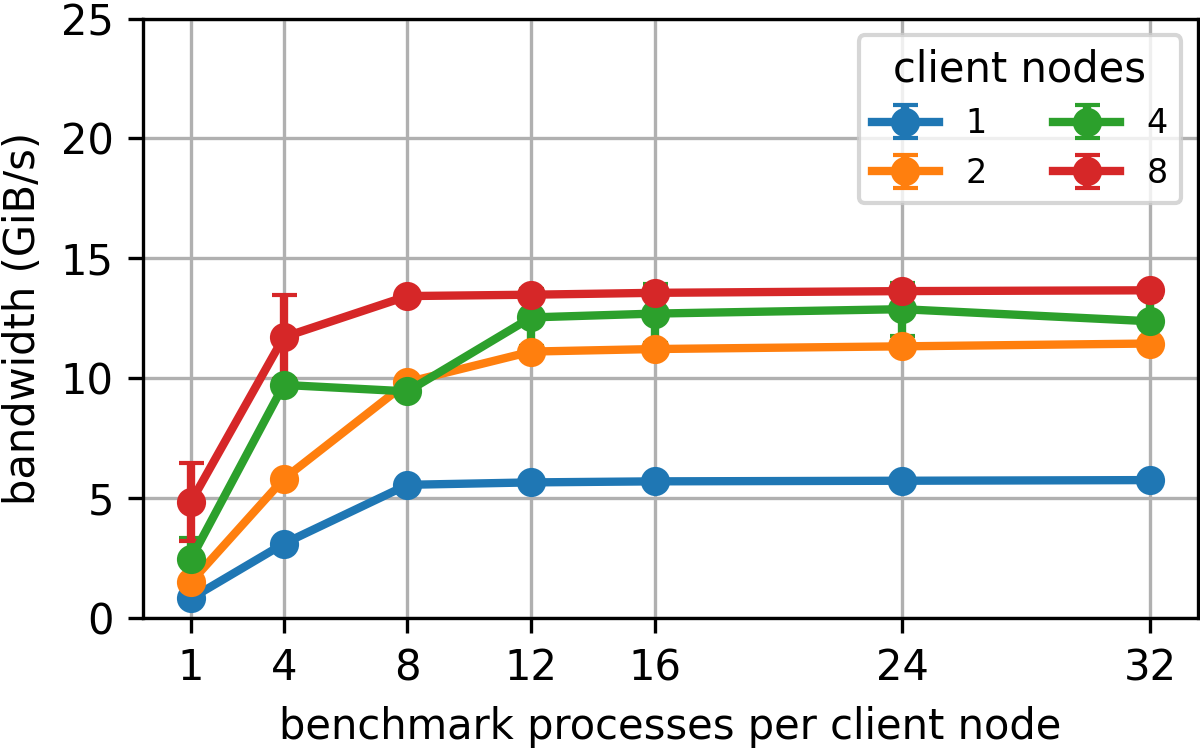}
        \caption{IOR/HDF5 on libdaos, Write}
    \end{subfigure}
    \begin{subfigure}[b]{112pt}
        \vspace{6pt}
        \includegraphics[width=112pt,trim={27pt 0 0 0},clip]{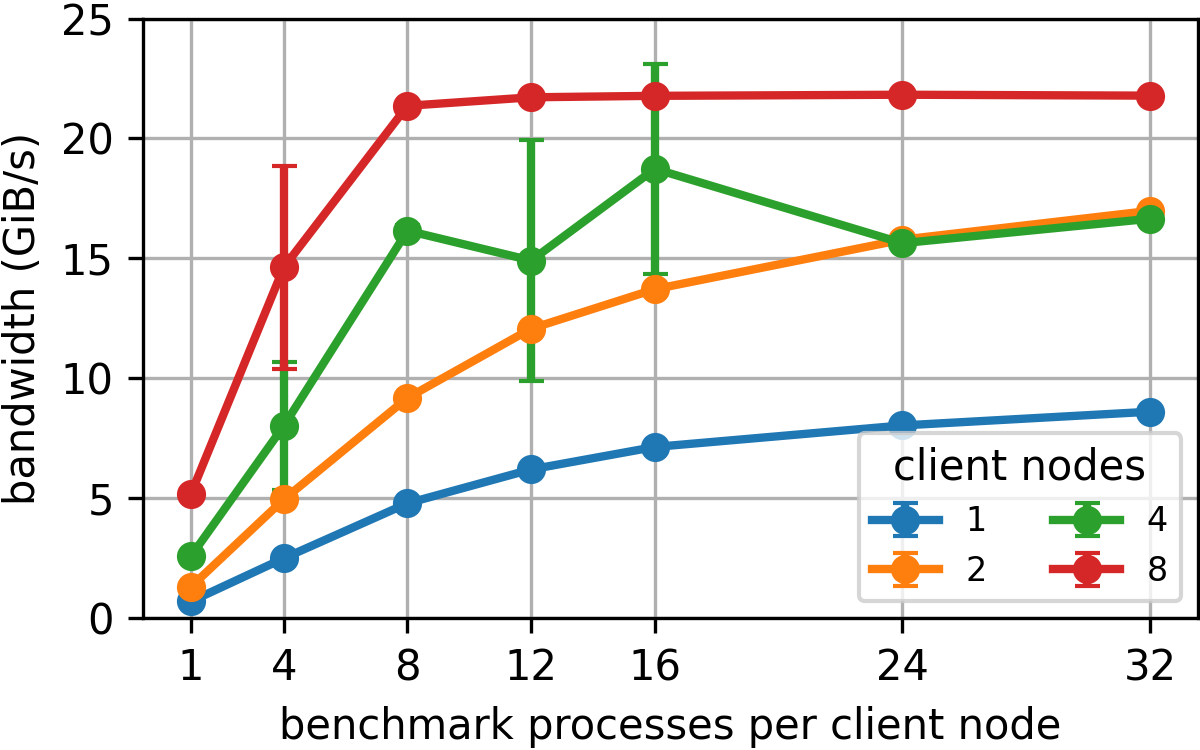}
        \caption{IOR/HDF5 on libdaos, Read}
    \end{subfigure}
    \caption{Client node and process count optimisation results for IOR on libdaos and IOR/HDF5 on libdaos, against a 4-node DAOS instance.}
    \label{fig:ior_hdf5_libdaos_4sn_cn_cpcn}
\end{figure}%

\begin{figure}[htbp]
    \centering
    \begin{subfigure}[b]{168pt}
        \centering
        \includegraphics[width=168pt,trim={0 28pt 0 0},clip]{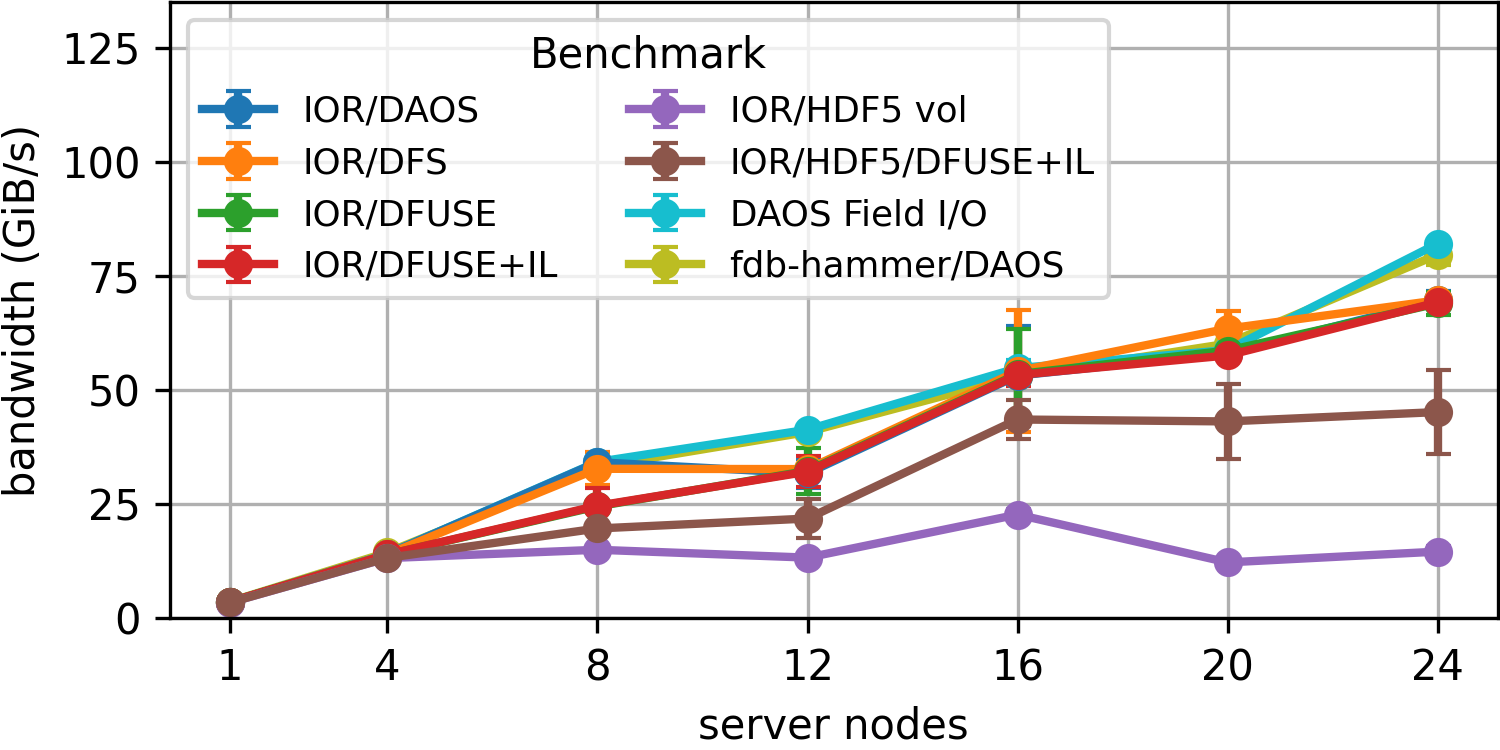}
    \end{subfigure}
    \begin{subfigure}[b]{168pt}
        \centering
        \vspace{6pt}
        \includegraphics[width=168pt,trim={0 0 0 0},clip]{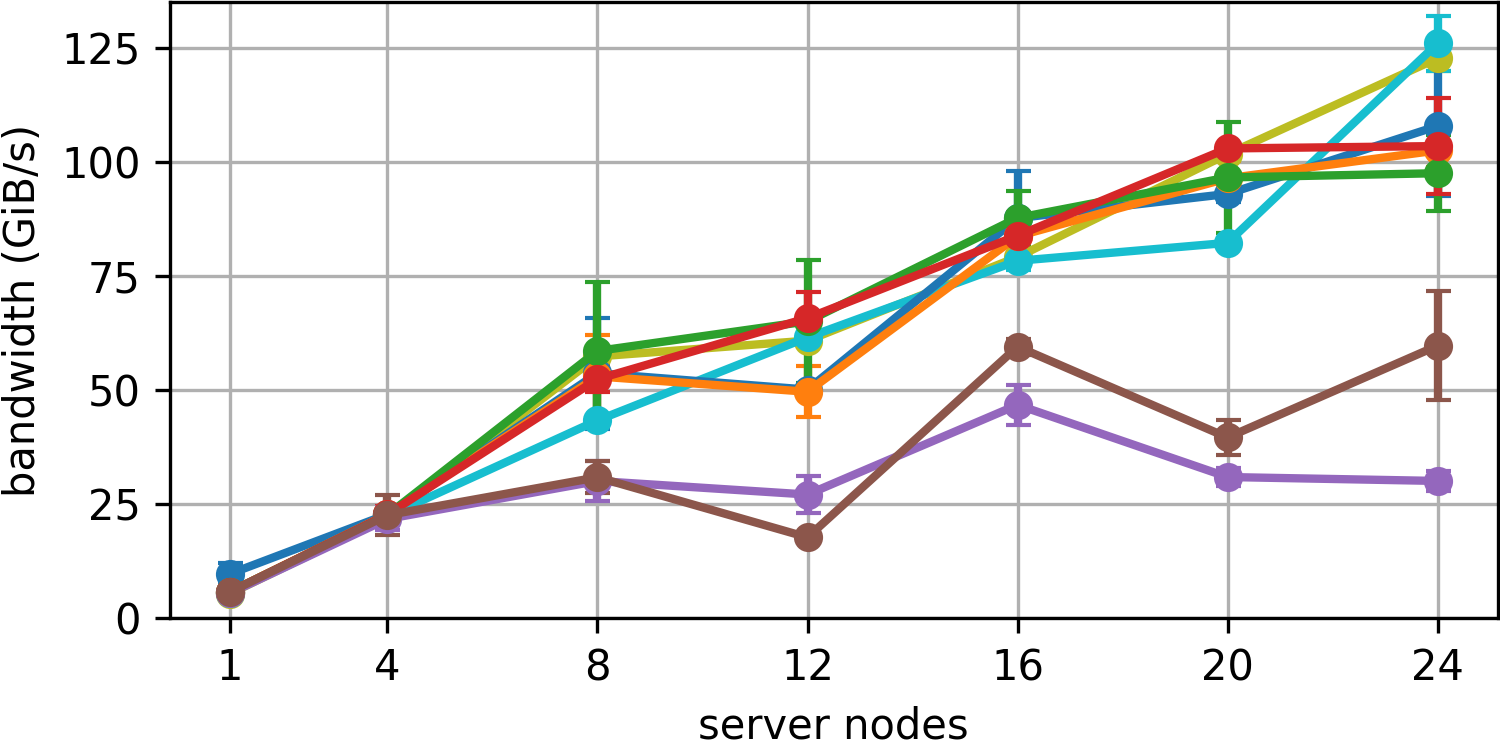}
    \end{subfigure}
    \caption{Write scalability (top) and read scalability (bottom) of the DAOS APIs and applications, with no redundancy.}
    \label{fig:scalability_no_redun}
\end{figure}%

HDF5 on DFUSE reaches approximately half the performance and seems to stop scaling beyond 16 server nodes, but it should be noted that this application was designed for POSIX and ported with zero effort or tuning other than the adjustment of the DFUSE file object class. HDF5 on libdaos performs worse and stops scaling beyond 4 DAOS server nodes, and this is likely due to the discussed container scalability issue.

Nevertheless, this comparison between HDF5 on DFUSE and HDF5 on libdaos makes an excellent example of the benefit of running POSIX applications on DFUSE directly. They may not achieve optimal performance, but they can reach substantial performance with zero porting effort, exceeding the performance that a full libdaos porting would provide if not following all DAOS best practices\cite{daos-best-practices}.

\subsection{Performance of DAOS redundancy}

We ran optimisation tests similar to those in Fig. \ref{fig:ior_16sn_cn_cpcn} and Fig. \ref{fig:apps_16sn_cn_cpcn}, against a 16-node DAOS system, this time using 2+1 erasure-coding, preserving the same sharding configuration as in previous tests. Some of the most relevant results for the erasure-code tests are shown in Fig. \ref{fig:ior_16sn_cn_cpcn_ec}.

\begin{figure}[htbp]
    \centering
    \begin{subfigure}[b]{124pt}
        \includegraphics[width=124pt,trim={0 16pt 0 0},clip]{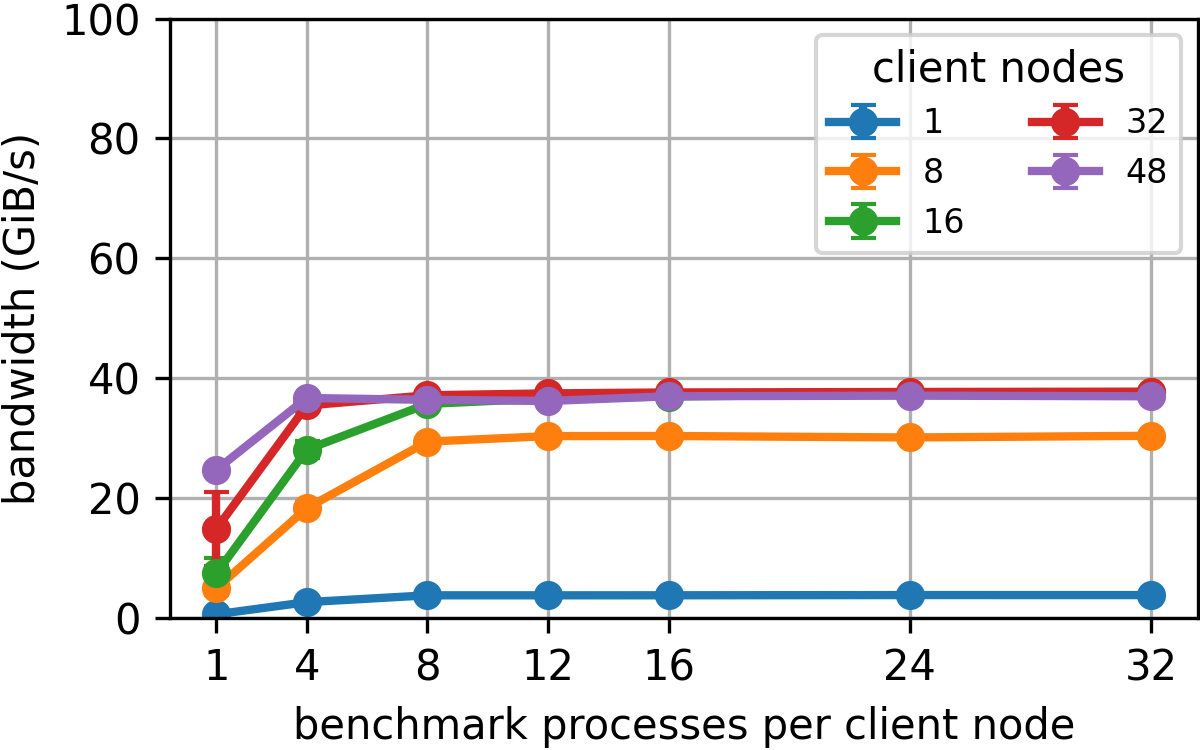}
        \caption{IOR on libdaos, Write}
    \end{subfigure}
    \begin{subfigure}[b]{109pt}
        \includegraphics[width=109pt,trim={35pt 16pt 0 0},clip]{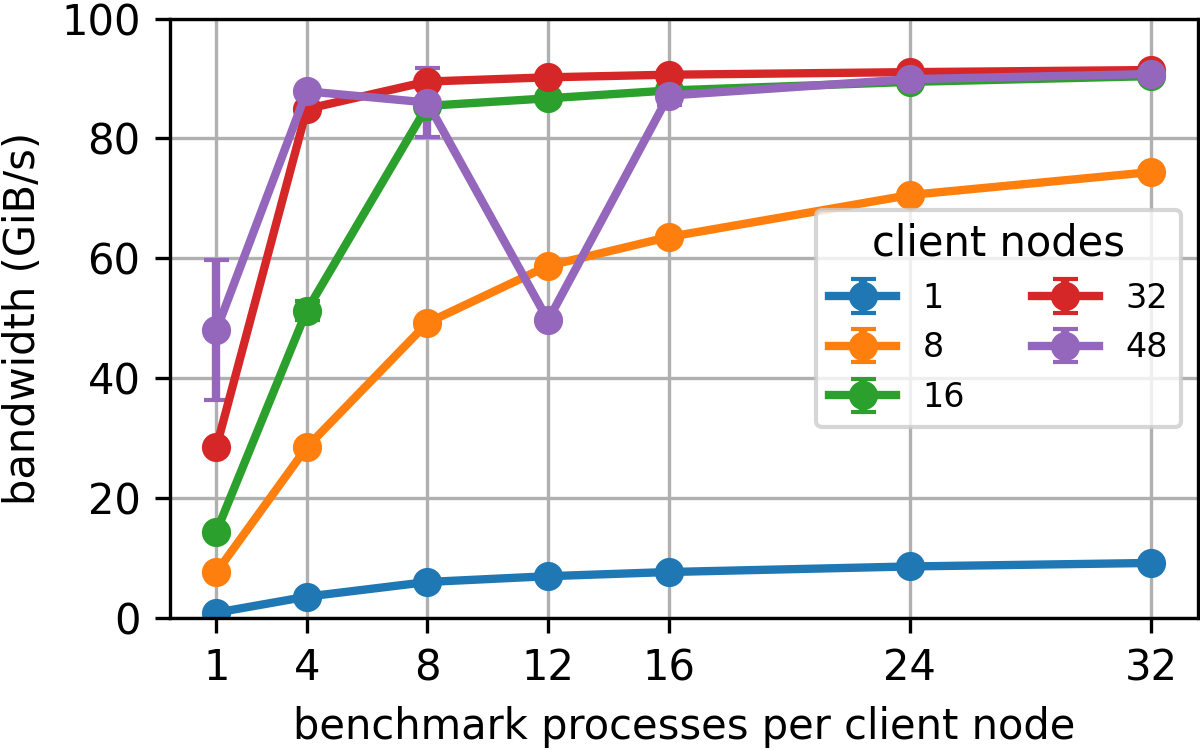}
        \caption{IOR on libdaos, Read}
    \end{subfigure}
    \begin{subfigure}[b]{124pt}
        \vspace{6pt}
        \includegraphics[width=124pt,trim={0 16pt 0 0},clip]{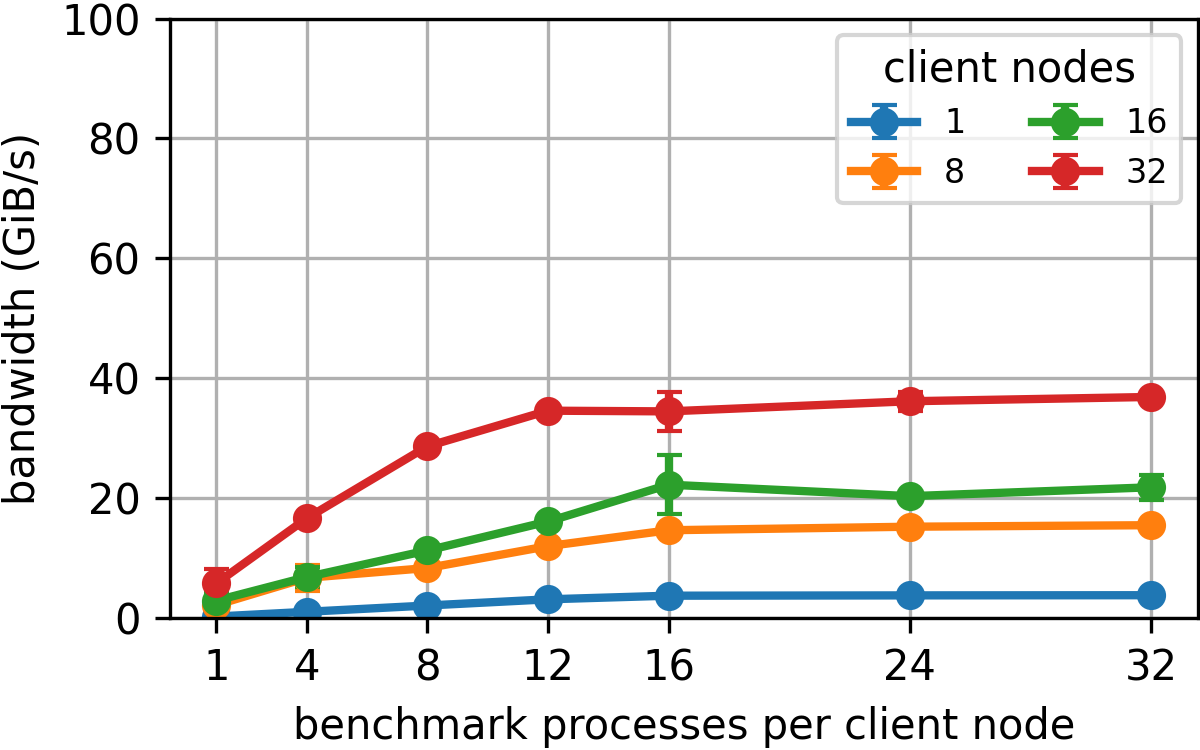}
        \caption{IOR on DFUSE+IL, Write}
    \end{subfigure}
    \begin{subfigure}[b]{109pt}
        \vspace{6pt}
        \includegraphics[width=109pt,trim={35pt 16pt 0 0},clip]{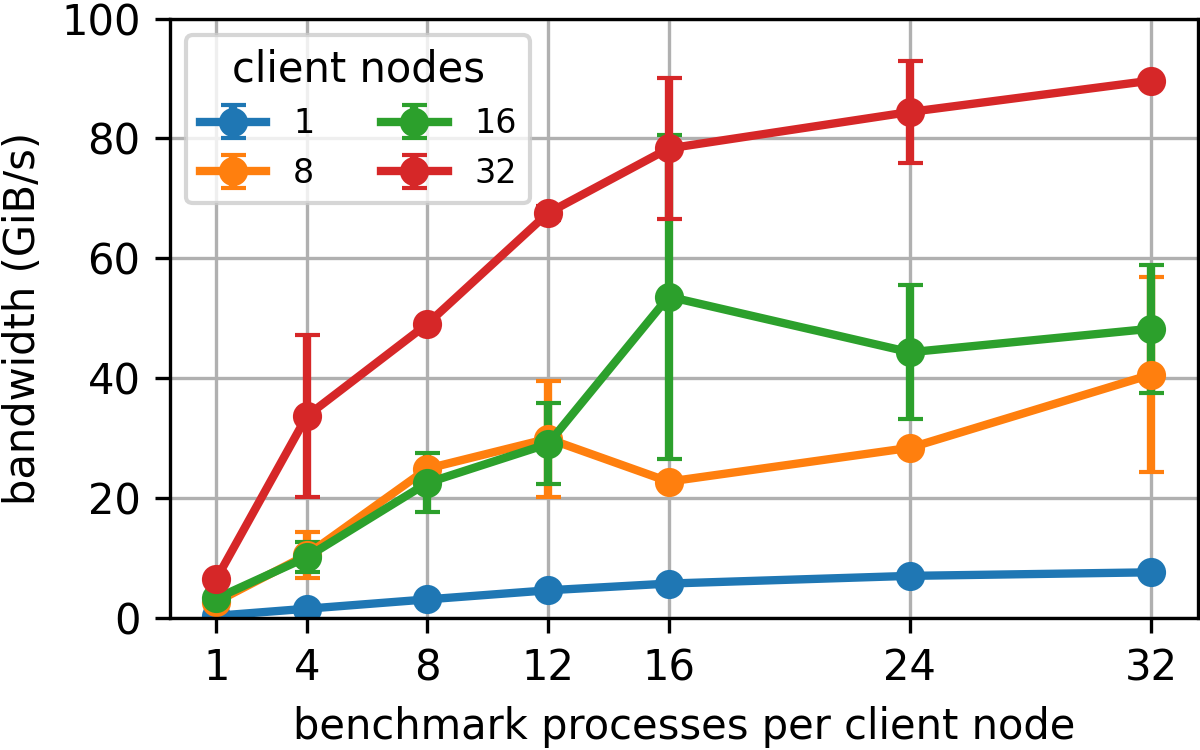}
        \caption{IOR on DFUSE+IL, Read}
    \end{subfigure}
    \begin{subfigure}[b]{124pt}
        \vspace{6pt}
        \includegraphics[width=124pt]{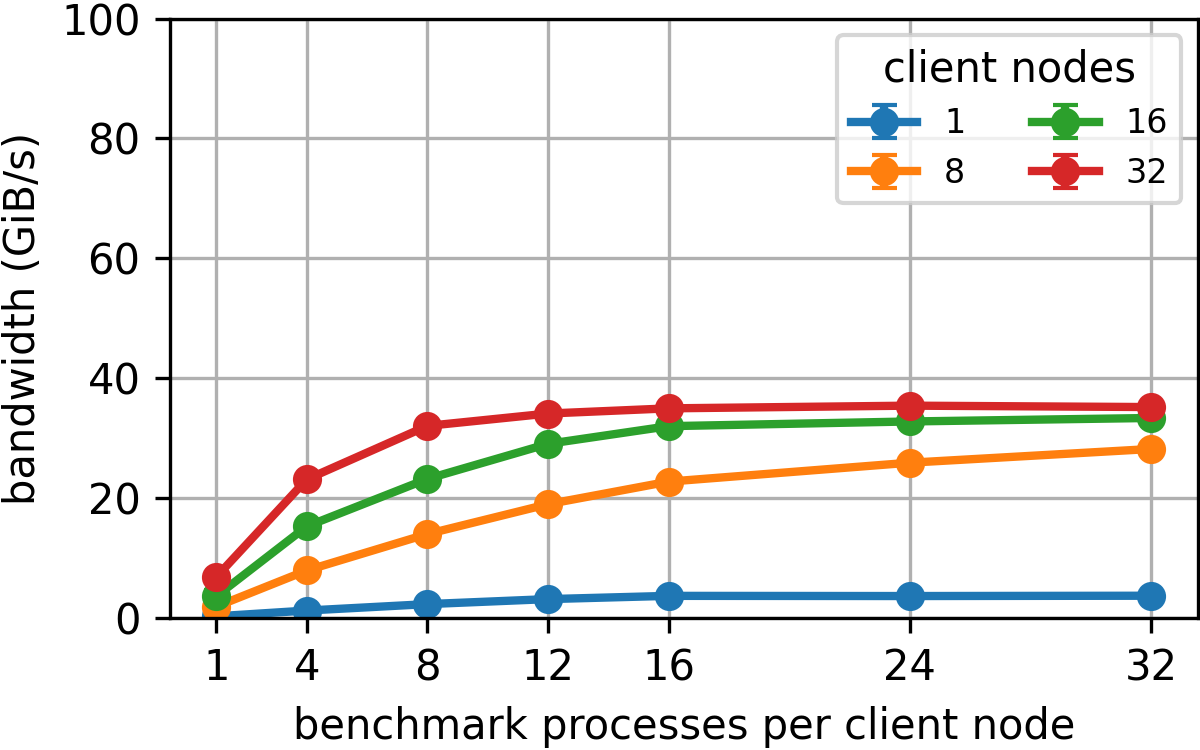}
        \caption{fdb-hammer on libdaos, Write}
    \end{subfigure}
    \begin{subfigure}[b]{109pt}
        \vspace{6pt}
        \includegraphics[width=109pt,trim={35pt 0 0 0},clip]{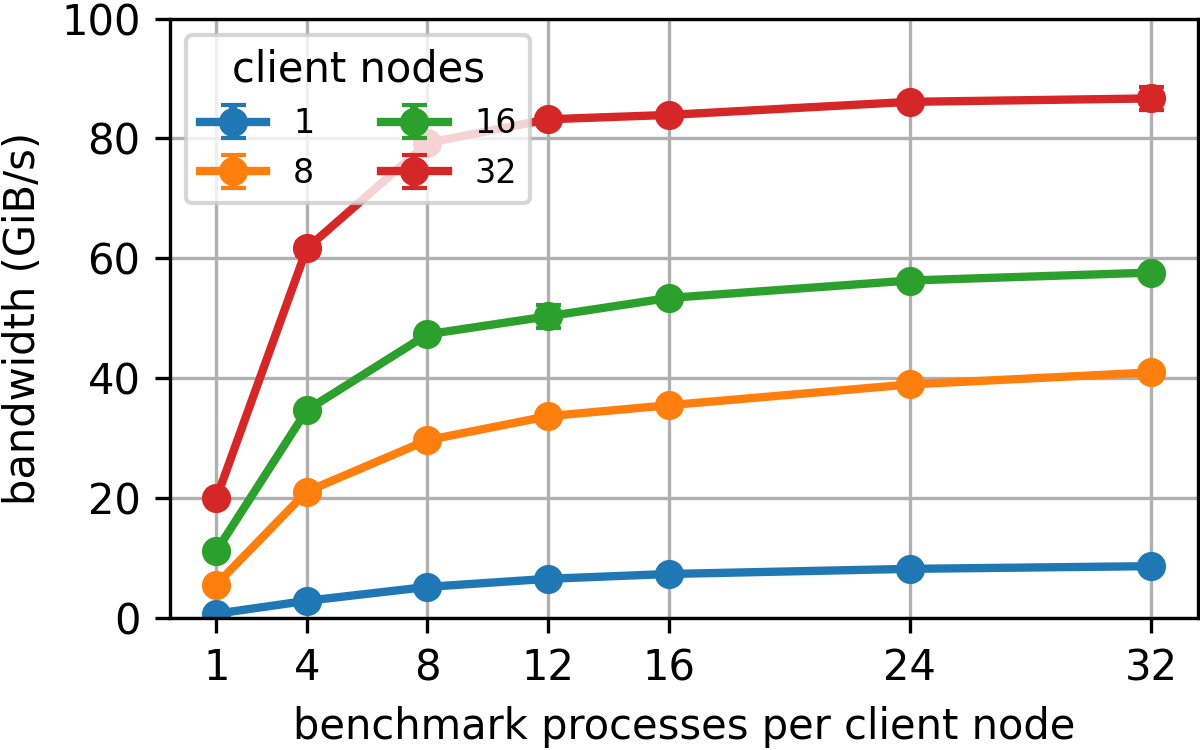}
        \caption{fdb-hammer on libdaos, Read}
    \end{subfigure}
    \caption{Client node and process count optimisation results for IOR and fdb-hammer against a 16-node DAOS instance with EC 2+1.}
    \label{fig:ior_16sn_cn_cpcn_ec}
\end{figure}

As expected, erasure-coding did not harm read performance, with all tested benchmarks reaching the same read bandwidths as before. For write, the maximum bandwidths achieved are close to 40 GiB/s, that is, two thirds of the bandwidths obtained with no redundancy enabled. This is essentially optimal relative to the hardware available, as for an erasure-code configuration of 2+1, an additional 50\% of data volume needs to be written, using 50\% more NVMe device bandwidth.

For these tests, directories in libdfs and Key-Values in Field I/O and fdb-hammer were configured with a replication factor of 2 rather than with 2+1 erasure-coding, as it can be inefficient to perform erasure-coding on indexing entities that are being constantly modified.

Similar tests with a replication factor of 2 for both the Arrays/files and Key-Values/directories, not included here, showed similarly optimal results, with read bandwidth unaffected, and write bandwidths halved reaching up to 30 GiB/s.

\subsection{Comparison to POSIX file system performance}

We deployed a Lustre distributed file system on 16 NVMe nodes with identical hardware to the DAOS deployment, having 16 Object Storage Targets (OSTs)\cite{lustre-ost} deployed on each, plus an additional node with a single NVMe drive where a Metadata Service (MDS)\cite{lustre-mds} was deployed. Lustre was configured with no data protection enabled. We ran IOR and the fdb-hammer benchmarks, configured to use their POSIX backends, against this Lustre system.

IOR results, not included here, showed Lustre can also reach close to optimal hardware performance for large file-per-process I/O. Results for fdb-hammer tests, also using a file per process with striping across 8 OSTs and a stripe size of 8 MiB, are shown in Fig. \ref{fig:fdbh_16sn_cn_cpcn_no_contention_lustre}.

These results show that fdb-hammer can reach close to IOR bandwidths for write, as fdb-hammer and its POSIX backend are optimised for write. fdb-hammer readers, however, only reach up to 40 GiB/s, and this is explained by the increased metadata workload, which Lustre and file systems in general are not optimised for. fdb-hammer on libdaos performing small I/O, however, can reach close to optimal performance for both write and read, as shown in Fig. \ref{fig:apps_16sn_cn_cpcn} (g) and (h).

\begin{figure}[htbp]
    \centering
    \begin{subfigure}[b]{124pt}
        \includegraphics[width=124pt]{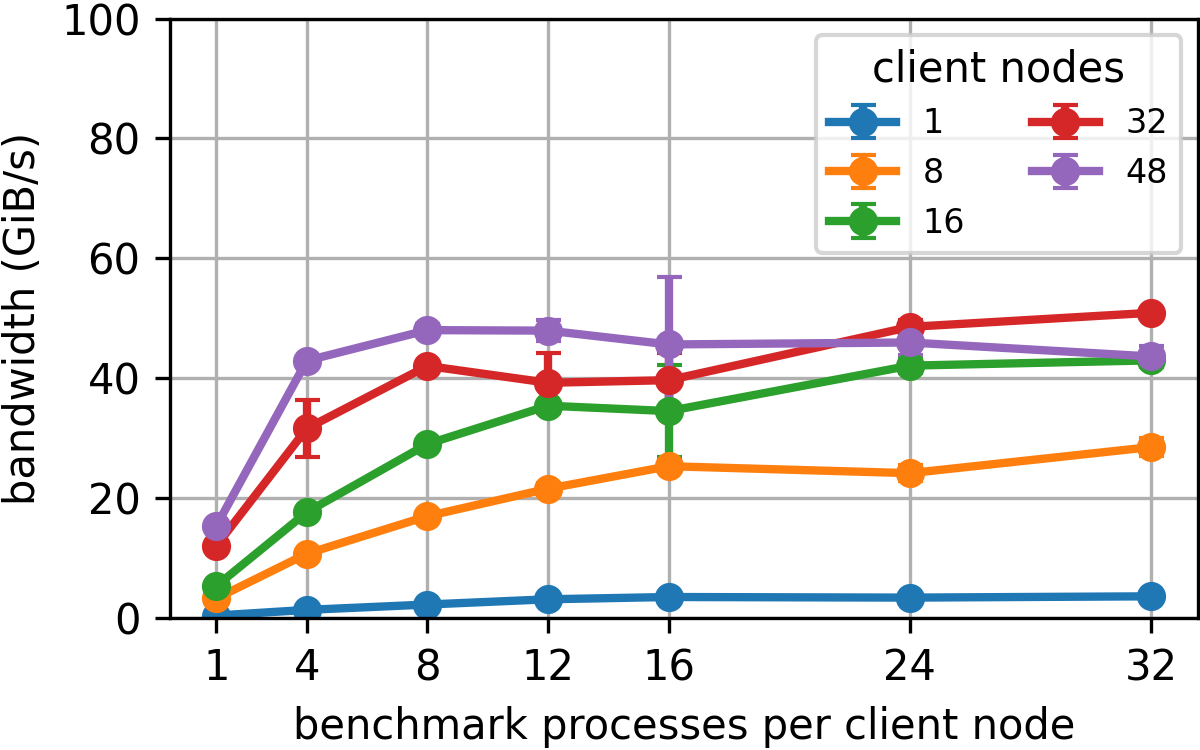}
        \caption{fdb-hammer on POSIX, Write}
    \end{subfigure}
    \begin{subfigure}[b]{109pt}
        \includegraphics[width=109pt,trim={35pt 0 0 0},clip]{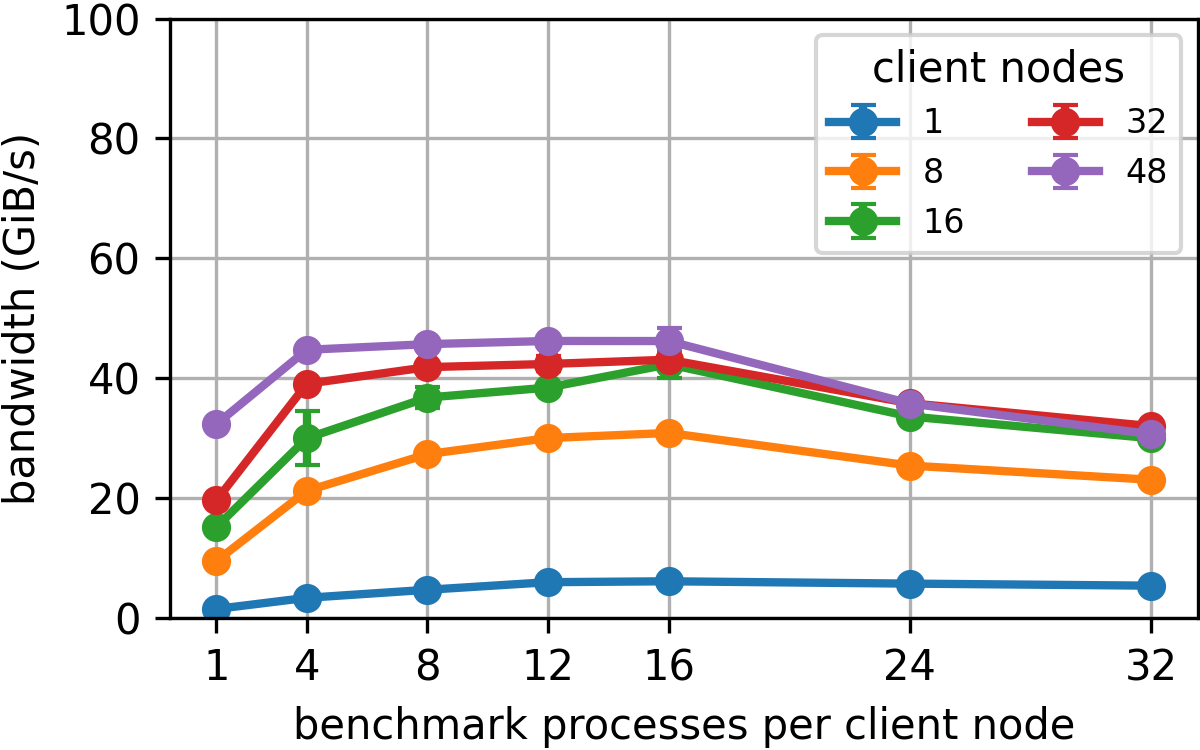}
        \caption{fdb-hammer on POSIX, Read}
    \end{subfigure}
    \caption{Client node and process count optimisation results for fdb-hammer runs on POSIX, against a 16+1-node Lustre instance.}
    \label{fig:fdbh_16sn_cn_cpcn_no_contention_lustre}
\end{figure}

\subsection{Comparison to Ceph object store performance}

Ceph\cite{ceph} --- another open-source object store designed for data protection on commodity hardware, popular in Cloud environments --- was deployed on 16 NVMe nodes, with 16 Object Storage Daemons (OSDs)\cite{ceph-osd} per node, plus an additional node with no NVMe where a Ceph Monitor\cite{ceph-mon} was deployed. Ceph was configured with no data protection. We ran IOR and fdb-hammer configured to use their Ceph backends, both making use of the librados library for direct operation with the core object storage functionality of Ceph.

IOR results using an object per process, not included here, showed Ceph could only reach up to 25 GiB/s for write and 50 GiB/s for read --- roughly a half of the corresponding results for DAOS and Lustre. This IOR performance difference is explained on one hand because Ceph cannot shard objects across OSDs unless enabling erasure-code or replication, whereas IOR on DAOS and Lustre did benefit from sharding even if data protection was disabled. On the other hand, we configured Ceph with the recommended maximum object size of 132 MiB, and adjusted the IOR processes to perform each only 100 I/O operations of 1 MiB to fit in the per-process objects. This prevented IOR processes sustaining I/O to their assigned object for very long, thus not saturating the system. Configuring Ceph for larger object sizes is discouraged and resulted in low write performance.

fdb-hammer runs on Ceph, for which results are shown in Fig. \ref{fig:fdbh_16sn_cn_cpcn_no_contention_ceph}, reached higher bandwidths of up to 40 GiB/s for write and 70 GiB/s for read. The number of Ceph Placement Groups (PGs)\cite{ceph-pg} was optimised, with the optimum value found to be 1024, to achieve balanced object placement across OSDs, and thus best performance. This performance increase with respect to IOR is explained by the fact that fdb-hammer processes perform 10k I/O operations of 1 MiB each, with a separate Ceph object for every I/O. This results in many objects being placed in a balanced way across PGs and thus efficiently exploiting all server bandwidth, and enables processes to sustain I/O for longer and saturate the system. The achieved performance, however, is only approximately two thirds of the ideal hardware bandwidth and significantly lower than the fdb-hammer bandwidths when run against DAOS. 

\begin{figure}[htbp]
    \centering
    \begin{subfigure}[b]{124pt}
        \includegraphics[width=124pt]{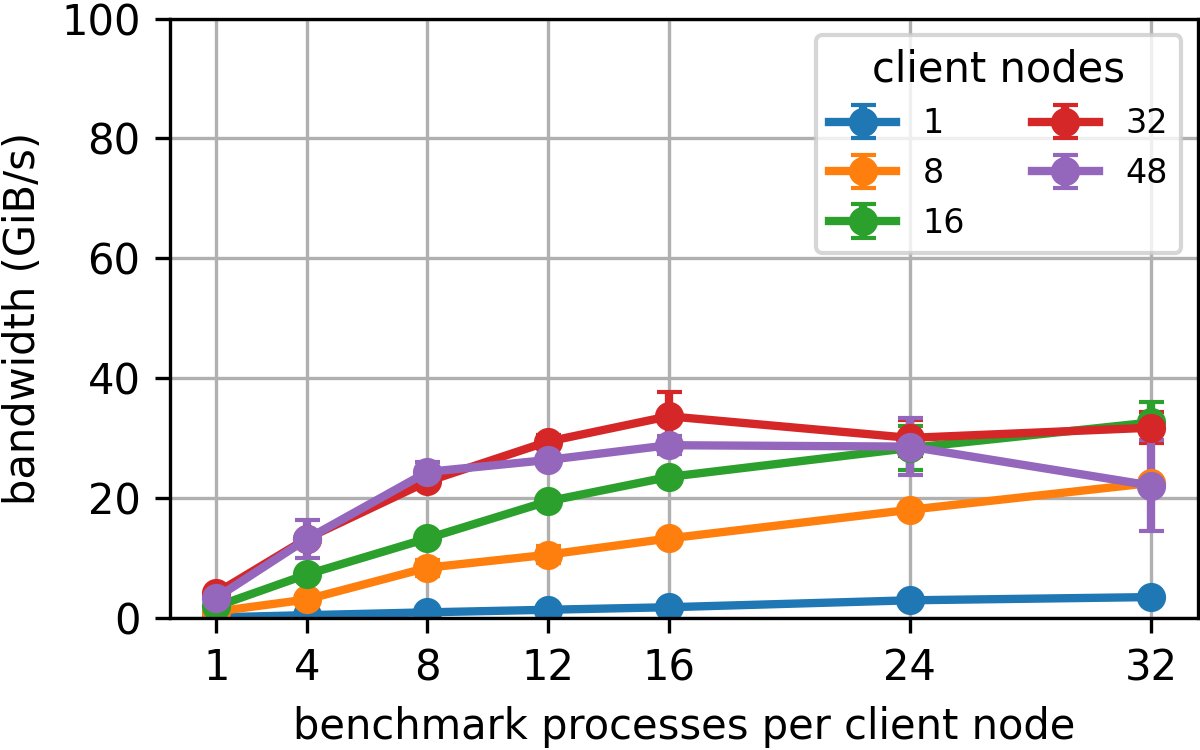}
        \caption{fdb-hammer on librados, Write}
    \end{subfigure}
    \begin{subfigure}[b]{109pt}
        \includegraphics[width=109pt,trim={35pt 0 0 0},clip]{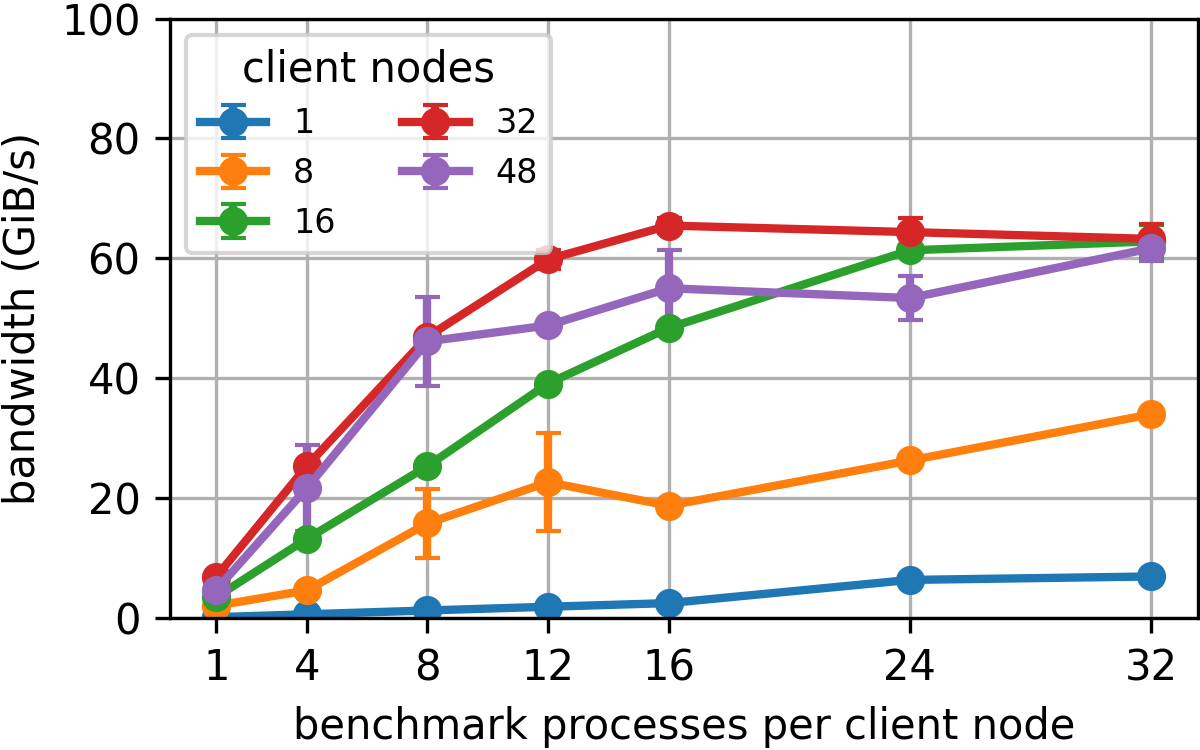}
        \caption{fdb-hammer on librados, Read}
    \end{subfigure}
    \caption{Client node and process count optimisation results for fdb-hammer runs on librados, against a 16+1-node Ceph instance.}
    \label{fig:fdbh_16sn_cn_cpcn_no_contention_ceph}
\end{figure}

These results indicate that Ceph can provide reasonable, albeit suboptimal, performance for applications using small object sizes, and is not designed to support high performance for large object sizes. These performance limitations could become more marked at larger scales.

Results for fdb-hammer runs on 32 client nodes against DAOS, Ceph and Lustre instances, extracted from Fig. \ref{fig:apps_16sn_cn_cpcn} (e) and (f), Fig. \ref{fig:fdbh_16sn_cn_cpcn_no_contention_lustre}, and Fig. \ref{fig:fdbh_16sn_cn_cpcn_no_contention_ceph}, are shown superimposed in Fig. \ref{fig:daos_vs_ceph_vs_lustre} for direct comparison of the reachable bandwidths with an application such as fdb-hammer, operating with small-sized data units, optimised for the three storage systems.

\begin{figure}[htbp]
    \centering
    \begin{subfigure}[b]{126pt}
        \includegraphics[width=126pt]{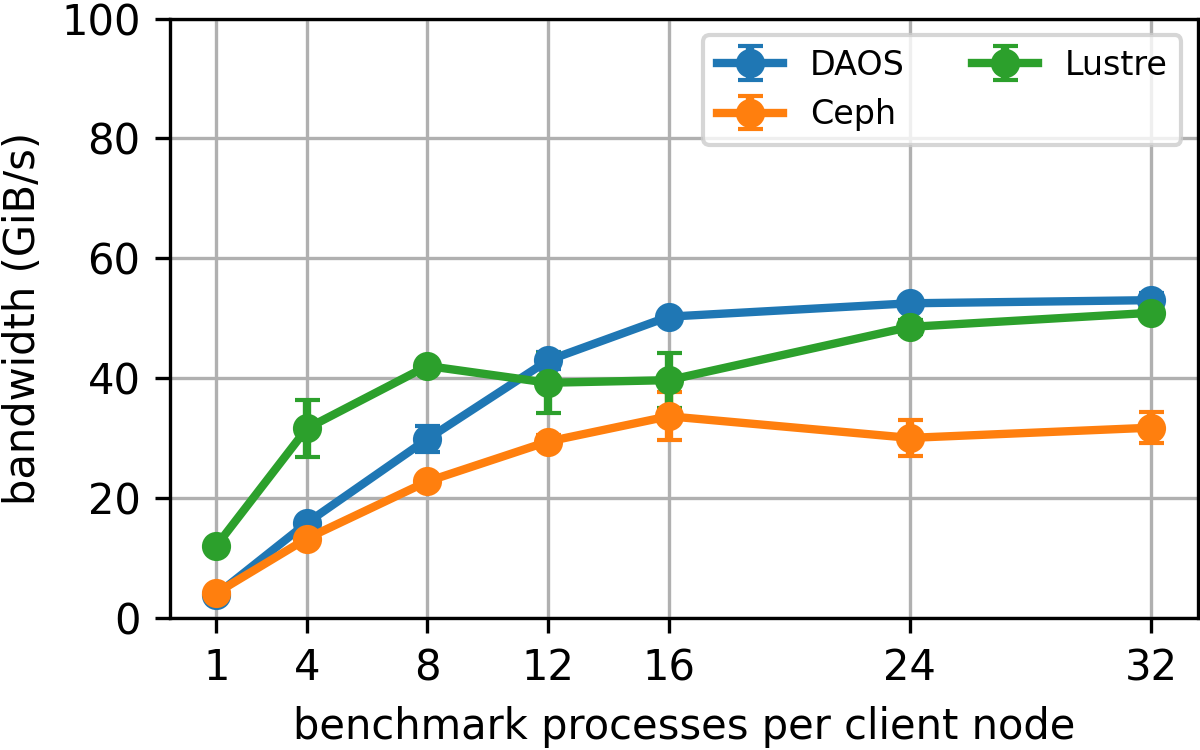}
        \caption{fdb-hammer, Write}
    \end{subfigure}
    \begin{subfigure}[b]{111pt}
        \includegraphics[width=111pt,trim={35pt -0.25pt 0 0},clip]{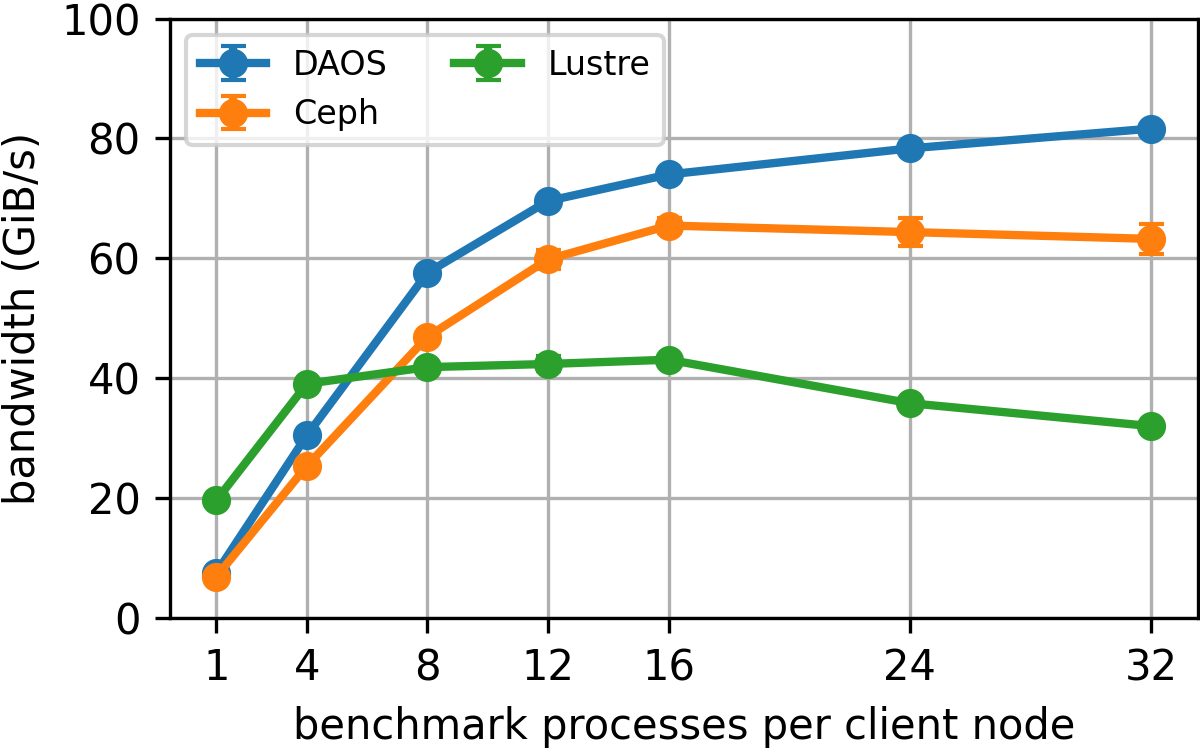}
        \caption{fdb-hammer, Read}
    \end{subfigure}
    \caption{Results for fdb-hammer runs on 32 client nodes against deployments of DAOS on 16 nodes, Ceph on 16+1 nodes, and Lustre on 16+1 nodes.}
    \label{fig:daos_vs_ceph_vs_lustre}
\end{figure}

\section{Conclusions}

This work has demonstrated that DAOS on NVMe and all DAOS interfaces perform and scale very well, at least up to the scale tested and with an I/O size of 1 MiB --- much smaller than any distributed file system could support while preserving high performance.

Applications making direct use of libdaos can perform and scale equally as well if following certain best practices for high performance. Even though HDF5 on DFUSE and HDF5's DAOS adaptor do not perform or scale as well as the rest, they have demonstrated that running existing POSIX applications on DFUSE can give reasonable performance with zero porting effort, and indeed better than porting to libdaos when best practices are not followed with that interface.

We have also demonstrated DAOS can support replication and erasure coding without degrading performance more than is strictly required given the underlying hardware, although it unavoidably implies a reduction in total storage capacity.

Finally, this work has briefly compared DAOS performance to Ceph and a distributed POSIX file system, and demonstrated the potential of DAOS as storage for HPC systems, as it is the only option that can provide high performance both for large I/O as well as for metadata and small I/O workloads. This can give more freedom for applications to perform I/O as desired without requiring a major redesign to adapt to specific well-performing I/O patterns.

\section*{Acknowledgment}

For the purpose of open access, the authors have applied a Creative Commons Attribution (CC BY) licence to any Author Accepted Manuscript version arising from this submission.

\twocolumn[%
{\begin{center}
\Huge
Appendix: Artifact Description/Artifact Evaluation        
\end{center}}
]


\appendixAD

\section{Overview of Contributions and Artifacts}

\subsection{Paper's Main Contributions}

\begin{description}
\item[$C_1$] Demonstration that DAOS on NVMe and all DAOS interfaces can perform and scale very well with an I/O size of 1 MiB.
\item[$C_2$] Demonstration that applications making direct use of libdaos can also scale very well, although HDF5 on libdaos or DFUSE does not scale so well.
\item[$C_3$] Demonstration that DAOS can support replication and erasure coding without degrading performance more than is strictly required.
\item[$C_4$] Demonstration that, compared to Lustre and Ceph, DAOS is the only option that can provide high performance both for large I/O as well as for metadata and small I/O workloads.
\end{description}

\subsection{Computational Artifacts}

\begin{description}
\item[$A_1$] https://doi.org/10.5281/zenodo.13757427
\end{description}

\begin{center}
\begin{tabular}{rll}
\toprule
Artifact ID  &  Contributions &  Related \\
             &  Supported     &  Paper Elements \\
\midrule
$A_1$   &  $C_1$, $C_2$, $C_3$, $C_4$ & All figures \\
\bottomrule
\end{tabular}
\end{center}

\vskip\baselineskip

\section{Artifact Identification}

\newartifact

\artrel

This artifact includes scripts and configuration, all contained in the \texttt{google} directory (all other directories are irrelevant to the analysis in this paper), to deploy Lustre and Ceph on NVMe on Google Cloud, as well as to deploy an auto-scaling Slurm cluster of client nodes to run benchmarks against DAOS, Lustre and Ceph deployments. Scripts and configuration to deploy DAOS systems are property of Google Cloud and have not been made publicly available.

The artifact also includes scripts and configuration to build and run the I/O benchmarks in the Slurm cluster to reproduce all tests reported in the paper.

Therefore, this artifact can reproduce all tests necessary to verify all contributions ($C_1$, $C_2$, $C_3$, $C_4$) of this paper.

\artexp

IOR on libdaos, libdfs, DFUSE and DFUSE+IL, and Field I/O and fdb-hammer against DAOS, should perform and scale very well up to 24 NVMe server nodes. IOR on HDF5 on libdaos and DFUSE+IL should perform and scale worse.

Using a DAOS replication factor of 2 or erasure-coding of 2+1 should not harm read performance, and should reduce write performance down to a half for replication and two thirds for erasure-coding.

IOR against Lustre should perform as well as IOR against DAOS deployed on equivalent amount of NVMe server nodes. fdb-hammer against Lustre should perform as well as against DAOS for write, and worse than DAOS for read.

IOR against Ceph should perform significantly worse than IOR against DAOS deployed on an equivalent amount of NVMe server nodes. fdb-hammer against Ceph should reach bandwidths approximately two thirds as high as those observed for fdb-hammer against DAOS.

\arttime

The Artifact Setup is expected to require 360 minutes, not including the time required to follow and manually execute the steps of the build procedure or to set up the Google Cloud account and quota. This is because a number of images need to be built for the different storage systems and the Slurm cluster, and also because the deployment of these complex systems takes a substantial amount of time until ready. Additionally, because the testing is performed on preemptible virtual machines, the deployment of these systems may need to be repeated several times before all tests are completed.

The Artifact Execution can require approximately 20000 minutes of wall-clock time. For every data point in every figure in the paper, three test repetitions need to be run, each requiring between 1 and 10 minutes of wall-clock time.

The artifact analysis can require approximately 1000 minutes of wall-clock time, as many log files need to be processed to calculate the bandwidth measurements, and the figures need to be generated.

\artin

\artinpart{Hardware}

The tests are run entirely in Google Cloud infrastructure, using virtual machines of type n2 and c2, some of them (the server machines) with locally attached NVMe SSDs. This artifact includes configuration resources which specify and provision the exact type of resources required.

\artinpart{Software}

The required software packages include DAOS, Ceph, Lustre, Slurm, MPI, IOR, HDF5 and its DAOS adaptor, Field I/O, and fdb-hammer. These --- except Slurm and MPI --- are introduced in the paper. The version numbers employed for the analysis have all been hard-coded in the build and install scripts in this artifact, which fetch all sources or binaries from public repositories.

\artinpart{Datasets / Inputs}

The benchmarks in this paper generally do not require special input files, as they generate random data to be written or read from the different storage systems. Where benchmarks require input configuration or seed data files, they have been included in this artifact.

\artinpart{Installation and Deployment}

As performed by the scripts included in the artifact.

\artcomp

The following steps need to be followed to execute this artifact.

\begin{itemize}

    \item Obtain a Google Cloud Platform account.

    \item Install Google Cloud SDK (last tested with v488.0.0), go (last tested with v1.22.1), Terraform (last tested with v1.9.4) and hpc-toolkit (last tested with v1.38.0).

    \item Clone this artifact.

    \item Run the scripts to provision the different storage systems (one at a time) and the Slurm client cluster, available in the artifact under \\
    \texttt{google/ceph/deployment/deploy.sh}, \texttt{google/lustre/deployment/deploy.sh}, and \texttt{google/slurm/deploy.sh}, respectively.

    \item Open an ssh connection to the Slurm controller node with \texttt{gcloud compute ssh}.
    
    \item Clone this artifact under \texttt{\$HOME/daos-tests} in the controller node.
    
    \item Locate the master test script, in function of the storage system and benchmark to be tested, available in the artifact under \\
    \texttt{google/<storage system>/<benchmark>/\-access\_patterns/A.sh}.

    \item Adjust the master test script with the amounts of server nodes (specified in the \texttt{servers} variable), client nodes (specified as a vector via the \texttt{C} variable) and processes per client node (specified as a vector via the \texttt{N} variable) to test with. The amount of I/O iterations per process and test repetitions can also be adjusted (via the variables \texttt{WR} and \texttt{REP}), but are generally set to 10k and 3 by default, respectively. For IOR tests, the IOR APIs to test with can also be adjusted via the \texttt{API} variable. For DAOS tests, the object class can be adjusted via the \texttt{OC} variable.

    \item Change directories to the \texttt{google/<storage system>} directory, and invoke the master script with \\
    \texttt{source <benchmark>/access\_patterns/A.sh}.

    \item When executed, a master script performs the following tasks:
    \begin{itemize}
        \item check that the storage system is available
        \item spin up the required Slurm client nodes
        \item build and/or install the client software if not present
        \item execute the benchmark in a loop for all configured client node and process counts, APIs, object classes, and repetitions
   \end{itemize}

    \item All test output is stored in a directory hierarchy under \texttt{google/<storage system>/runs}.
    
\end{itemize}

\artout

For every test run, the timestamp reported before the first I/O and the timestamp reported after the last I/O across all parallel processes have to be extracted from the log files for that run, and the time difference between the two calculated. The total amount of data written or read by all processes in the test run is then divided by the time in seconds obtained in the previous step, resulting in a bandwidth measurement in bytes per second. The average and standard deviation of the bandwidths for the three repetitions for every test are calculated and used as data point for the figures.


\begin{thebibliography}{00}
\bibitem{daos-scfa2022} Z. Liang, J. Lombardi, M. Chaarawi, and M. Hennecke, "DAOS: A Scale-Out High Performance Storage Stack for Storage Class Memory", In: Panda, D. (eds) Supercomputing Frontiers. SCFA 2020. Lecture Notes in Computer Science(), vol 12082. Springer, Cham. DOI:10.1007/978-3-030-48842-0\_3.
\bibitem{io500-sc23} A. Dilger, D. Hildebrand, J. Kunkel, J. Lofstead, G. Markomanolis, S. Ihara, and H. Nolte, "IO500 10 node list Supercomputing 2023", November 2023. https://io500.org/list/sc23/ten-production
\bibitem{io-contention-filesystems} A. K. Paul, O. Faaland, A. Moody, E. Gonsiorowski, K. Mohror, and A. R. Butt, "Understanding HPC Application I/O Behavior Using System Level Statistics", 2020 IEEE 27th International Conference on High Performance Computing, Data, and Analytics (HiPC), Pune, India, 2020, pp. 202-211, doi: 10.1109/HiPC50609.2020.00034.
\bibitem{lockwood} G. Lockwood, "What's so bad about POSIX I/O?". The Next Platform 2017. https://www.nextplatform.com/2017/09/11/whats-bad-posix-io/
\bibitem{lustre-internals} A. George, and R. Mohr, "Understanding Lustre Internals", 2024.\\https://wiki.lustre.org/Understanding\_Lustre\_Internals
\bibitem{gpfs-internals} F. Schmuck, and R. Haskin, "GPFS: A Shared-Disk File System for Large Computing Clusters", 2002.\\https://www.usenix.org/legacy/publications/library/proceedings/fast02/
full\_papers/schmuck/schmuck\_html/index.html
\bibitem{dfuse} "DAOS File System", 2024.\\https://docs.daos.io/v2.4/user/filesystem/\#dfuse-daos-fuse
\bibitem{daos-ipdps} N. Manubens, T. Quintino, S. D. Smart, E. Danovaro, and A. Jackson, "DAOS as HPC Storage: a View From Numerical Weather Prediction", 2023 IEEE International Parallel and Distributed Processing Symposium (IPDPS), St. Petersburg, FL, USA, 2023, pp. 1029-1040, doi: 10.1109/IPDPS54959.2023.00106.
\bibitem{daos-lustre-pasc} N. Manubens, S. D. Smart, E. Danovaro, T. Quintino, and A. Jackson, "Reducing the Impact of I/O Contention in Numerical Weather Prediction Workflows at Scale Using DAOS", 2024 Proceedings of the Platform for Advanced Scientific Computing Conference (PASC '24). Association for Computing Machinery, New York, NY, USA, Article 12, 1–12. https://doi.org/10.1145/3659914.3659926
\bibitem{ior} "HPC IO Benchmark Repository", 2024, GitHub repository.\\https://github.com/hpc/ior
\bibitem{hdf5} The HDF Group. "Hierarchical Data Format, version 5 [Computer software]", https://github.com/HDFGroup/hdf5
\bibitem{hdf5-daos-vol} J. Soumagne et al., "Accelerating HDF5 I/O for Exascale Using DAOS," in IEEE Transactions on Parallel and Distributed Systems, vol. 33, no. 4, pp. 903-914, 1 April 2022, doi: 10.1109/TPDS.2021.3097884
\bibitem{field-io} N. Manubens, A. Jackson, T. Quintino, S. Smart and E. Danovaro, "DAOS weather field I/O tests", 2024, GitHub repository ecmwf-projects/daos-tests (0.2.0). DOI:10.5281/zenodo.10699254.
\bibitem{fdb-hammer} "fdb-hammer.cc", 2024, GitHub repository file.\\https://github.com/ecmwf/fdb/blob/master/src/fdb5/tools/fdb-hammer.cc
\bibitem{fdb-pasc2019} S. Smart, T. Quintino, and B. Raoult. "A High-Performance Distributed Object-Store for Exascale Numerical Weather Prediction and Climate". In Proceedings of the Platform for Advanced Scientific Computing Conference (PASC '19). Association for Computing Machinery, New York, NY, USA, Article 16, 1–11. DOI:10.1145/3324989.3325726.
\bibitem{gcp-instance-types} "General-purpose machine family for Compute Engine", 2024.\\https://cloud.google.com/compute/docs/general-purpose-machines
\bibitem{gcp-smt} "Set the number of threads per core", 2024.\\https://cloud.google.com/compute/docs/instances/set-threads-per-core
\bibitem{daos-storage-model} "DAOS Storage Model", 2024.\\https://docs.daos.io/v2.0/overview/storage/
\bibitem{daos-best-practices} A. Jackson and N. Manubens, "Profiling and identifying bottlenecks in DAOS", DAOS User Group 2023 \\https://www.research.ed.ac.uk/files/392151952/DUG23\_FDB\_DAOS.pdf
\bibitem{lustre-ost} "Lustre Object Storage Service (OSS)", 2024.\\https://wiki.lustre.org/Lustre\_Object\_Storage\_Service\_(OSS)
\bibitem{lustre-mds} "Lustre Metadata Service (MDS)", 2024.\\https://wiki.lustre.org/Lustre\_Metadata\_Service\_(MDS)
\bibitem{ceph} "Welcome To Ceph", 2024.\\https://docs.ceph.com/en/reef/
\bibitem{ceph-osd} "Ceph-OSD -- Ceph Object Storage Daemon", 2024.\\https://docs.ceph.com/en/latest/man/8/ceph-osd/
\bibitem{ceph-mon} "Ceph-Mon -- Ceph Monitor Daemon", 2024.\\https://docs.ceph.com/en/latest/man/8/ceph-mon/
\bibitem{ceph-pg} "Placement Groups", 2024.\\https://docs.ceph.com/en/latest/rados/operations/placement-groups/
\end{thebibliography}
\end{document}